\documentclass[11pt]{amsart}

\usepackage{amsmath,mathtools}
\usepackage[dvipsnames,svgnames]{xcolor}
\usepackage{amsmath, amsthm, amssymb, amsfonts, enumerate}
\usepackage[colorlinks=true,linkcolor=blue,urlcolor=blue]{hyperref}
\usepackage{dsfont}
\usepackage{color}
\usepackage{geometry}
\usepackage [latin1]{inputenc}
\usepackage{comment}
\usepackage{subcaption}
\usepackage{graphicx}
\usepackage{epstopdf}
\usepackage{bbm}
\usepackage{geometry}

\usepackage{amsfonts}
\usepackage{amsfonts}
\usepackage{textcomp}

\usepackage{amssymb}
\usepackage{float}
\usepackage{tikz}
\usepackage{epsfig}
\usepackage{amsmath}
\usepackage[english]{babel}
\usepackage{a4}
\usepackage{enumerate}
\usepackage{tabularx}
\usepackage{arydshln}
\usepackage{array}

\usepackage{csquotes}
\usepackage{multirow}
\usepackage{setspace}

\newcommand{\one}{\text{$\mathbbm{1}$}}

\renewcommand{\tilde}{\widetilde}

\renewcommand{\epsilon}{\varepsilon}

\newtheorem{theorem}{Theorem}[section]
\newtheorem{remark}[theorem]{Remark}
\newtheorem{assumption}[theorem]{Assumption}

\newtheorem{proposition}[theorem]{Proposition}

\newtheorem{definition}[theorem]{Definition}

\usepackage[
natbib=true,
backend=biber,
style=authoryear-comp,
sorting=nyt,
maxcitenames = 2,
maxbibnames=10,
giveninits=true,
uniquelist=false,
uniquename=init
]{biblatex}

\AtEndDocument{%
  \vspace{1cm}
  \begin{tabular}{@{}l@{}}%
    \noindent\textsc{Felix Dammann:}\\
    \noindent\textsc{Center for Mathematical Economics (IMW), Bielefeld University, Germany}\\
    \noindent\textit{Email address}: \texttt{dammann@uni-bielefeld.de}
  \end{tabular} \par 
  \vspace{0.3cm}
  \begin{tabular}{@{}l@{}}%
    \textsc{Giorgio Ferrari:} \\
    \textsc{Center for Mathematical Economics (IMW), Bielefeld University, Germany}\\
    \textit{Email address}: \texttt{giorgio.ferrari@uni-bielefeld.de}
  \end{tabular}}

\title[Green Technology Adoption with Endogenous Carbon Price]{A Stationary Equilibrium Model of Green Technology Adoption with Endogenous Carbon Price}

\author{Felix Dammann}
\author{Giorgio Ferrari}

\date{\today}

\numberwithin{equation}{section}

\begin{document}

\begin{abstract}
This paper proposes and analyzes a stationary equilibrium model for a competitive industry which endogenously determines the carbon price necessary to achieve a given emission target. In the model, firms are identified by their level of technology and make production, entry, and abatement decisions. Polluting firms are subject to a carbon price and abatement is formulated as an irreversible investment, which entails a sunk cost and results in the firms switching to a carbon neutral technology. In equilibrium, we identify a carbon price and a stationary distribution of incumbent, polluting firms, that guarantee the compliance with a certain emission target. Our general theoretical framework is complemented with a case study with Brownian technology shocks, in which we discuss some implications of our model. We observe that a carbon pricing system alongside installation subsidies and tax benefits for green firms trigger earlier investment, while higher income taxes for polluting firms may be distorting. Moreover, we discuss the role of a welfare maximizing regulator, who, by optimally setting the emission target, may mitigate or revert some parameters' effects observed in the model with fixed limit. 
\end{abstract}

\maketitle

\section{Introduction}

We develop a model of carbon pricing within the framework of a general equilibrium model with strategic green technology adoption.
The analysis is conducted for a competitive economy that is in a steady state or stationary equilibrium, where equilibrium variables remain constant over time. \par 

Climate change is one of the major topics in today's world and addressing the challenge of reducing global emissions, as a large contributor of global warming, is a broadly studied and thoroughly discussed topic across several disciplines.  
Within the economic literature, ever since the contributions of \citet{Nordhaus1977}, it is widely accepted that any meaningful and efficient climate change policy to decrease emissions has to impose a price on the emissions of carbon dioxide and other greenhouse gases (cf.~\cite{Stern2007}). Its underlying idea is that by attaching a price to emissions, it creates financial incentives for those regulated to reduce their emissions, encourage them to adopt cleaner technologies, or to invest in renewable energy sources.  As of today, approximately 23\% of the world's total carbon emissions are subject to a carbon pricing system designed with the explicit goal of constraining emissions (cf.~\cite{WorldBank2023}).

The study and development of (effective) mechanisms for attaching a price to emission of pollutants is an important topic in the literature on environmental economics, and essentially boils down to two fundamental concepts.  

In the market-based or quantity approach, a regulator determines the maximum allowable level of overall emissions. Often implemented via a so-called cap-and-trade mechanism, it is considered to be one of the most promising and cost-effective market mechanisms in the effort to reduce carbon emissions. A prominent example includes the European Union's emission trading system, which was implemented in $2005$ and includes around $10.000$ installations that cover around $40\%$ of the total emissions. By setting a legally binding target for the maximum allowed level of greenhouse gas (GHG) emissions, this approach allows regulators to commit to concrete emission reduction goals on international, national or even industry specific level. Emission allowances are distributed to firms, either through auctions or as free allocations, and firms buy and trade allowances among themselves in order to account for their emissions. Consequently, this market-driven approach naturally gives rise to a market price for emission permits. \\
In contrast to this mechanism, where the quantity of overall emissions is limited, the price approach to carbon emissions directly links the emission of one unit of pollutant to a fixed cost. Typically imposed by a regulator through a tax or penalty, the resulting level of overall emissions is a priori unclear, as market participants are not regulated in the amount of emissions they are allowed to emit. However, the carbon price may be determined by estimates of the price required to limit emissions below some predetermined level. 

In both cases, a carbon price -- also referred to as \textit{carbon tax} or \textit{social cost of carbon} -- emerges. The price encapsulates the cost associated with the right to emit one unit of pollutant, thus serving the purpose of correcting market distortions that result from market participants not weighing in the external effects of their harmful activities. As emphasized by \citet{Nordhaus2007}, ``the key economic issue is how to balance the benefits and costs of global emissions reductions." Moreover, a transparent and comparable carbon price is essential to provide incentives to firms as well as stimulate research and development in carbon neutral technologies \citep[cf.][]{Stern2007, Zhao2003}. In particular, it should transmit the social cost of carbon emissions to the decisions of firms as well as individuals. \par
In the literature, there exist broad contributions addressing the topic of carbon pricing and the economic impact of climate change in various climate economics settings. \citet{Nordhaus2014} discussed the concept of social cost of carbon in order to mitigate losses that are caused by carbon emissions. \citet{Golosov2014} study a dynamic stochastic general equilibrium model, establishing a flexible optimal tax formula compatible with the climate impacts considered in \citet{Nordhaus2014}. \citet{Acemoglu2012} and \citet{Acemoglu2016} analyse directed technical change as well as optimal carbon taxes and subsidies. Other studies focused on the price formation of emission allowances, including works by \citet{Grull2011} as well as \citet{Carmona2009, Carmona2010, Carmona2011}, who establish market equilibria as well as design optimal emission trading mechanisms. The effect of cap-and-trade mechanisms and carbon taxes on firms' production and abatement decisions was studied in \citet{Anand2020, Fan2023}.
We also mention \citet{Aid2023}, proposing a Stackelberg-type equilibrium model with a regulator optimally offsetting shocks in the carbon price dynamics; \citet{Colla2012}, who examine endogeneous prices of permits and study the social welfare optimizing policy of a regulator; \citet{DeAngelis2023Climate} studying the effect of green investors on firms' abatement strategies; \citet{Barnett2023}, who analyses the impact of climate change uncertainty on economic and financial outcomes, as well as \citet{Hitzemann2018}, studying trading rounds of a finite set of firms which maximize over abatement and trading. 
\par

In our model, we seek to embed the issue of carbon price formation into a stationary equilibrium model with strategic abatement investment. Firms are characterized by their idiosyncratic technology shocks, which evolve according to a general It\^{o}-diffusion process, and we assume that firms' activities lead to carbon emissions. Since emissions are regulated and firms' profits are lowered through the implementation of a carbon price, they have the incentive to invest in a carbon-neutral technology. In equilibrium, we determine a stationary distribution of incumbent, polluting firms, the entry rate of new firms, as well as an optimal investment rule that is triggered by an equilibrium carbon price. The latter is endogenously determined and precisely reflects the cost necessary to achieve a certain emission target. Under reasonable and quite general assumptions on the involved functions and diffusion processes we are able to prove existence and uniqueness of this equilibrium.  

For exposition, in Section \ref{ECP:Section:ExplicitModel}, we establish a specific formulation of the problem. We consider the case of technology shock processes evolving according to arithmetic Brownian motions and an AK-structure for firms' production functions. Additionally, we assume a damage function of emissions on production, that admits a similar structure as the one considered in \citet{Golosov2014}. 
Within this particular model, we conduct a comparative statics analysis to explore some of the parameters' effects on the equilibrium values. For instance, we study the effect of a shift in the corporate tax rates of polluting or carbon neutral firms. Potential regulatory strategies might involve installation subsidies or different tax rates for these installations to incentivise polluting firms to invest in a carbon-neutral technology (see, for example, \cite{Farzin2000}). We find that a policy penalizing polluting firms through increased taxes may yield suboptimal results, since firms' reduced profits discourage potential firms from entering and thus causes a decreased market competition. This, in turn, leads to a smaller force for firms to become carbon-neutral, and we observe a decreasing carbon price that results in delayed investment. On the other hand, tax benefits for carbon neutral firms as well as subsidies on the investment cost lead to an opposing effect. Indeed, we observe that improving market conditions for carbon neutral firms prompts earlier investment by polluting firms and leads to a slightly higher equilibrium carbon price. 
Furthermore, we observe that reducing the cumulative emissions available to firms intensifies competition among them, which leads to an increased equilibrium carbon price. This effect also prompts firms to adopt a carbon-neutral technology earlier.

It is important to notice that cumulative emissions are not part of the equilibrium values at this point, but are exogenously given (for discussions on this, we refer to \citet{Stern2007, Nordhaus2014}). Clearly, this limit may be determined based on environmental considerations, such as the maximal allowed amount of emissions that can be released into the atmosphere while achieving a certain carbon concentration or limiting the global temperature increase to a certain degree. As such, the question of optimally setting the emission cap is far from being trivial. In the considered case study, we explore the concept of a regulator aiming to maximize social welfare by optimally setting the emission cap. Inspired by related contributions (see, for example, \cite{Ulph1996, Colla2012, Aid2023}), we investigate a regulator balancing the benefits from emission inducing production and their societal damages. Our comparative statics analysis also involves a comparison of the equilibrium values with exogenous and endogenous emission limits.  We observe that the regulator's actions potentially counteract or mitigate effects on carbon price and investment decisions observed in the analysis with fixed overall emissions. For instance, an increased carbon intensity of production or elevated emission induced damages on production lead the regulator to set a lower emission target, encouraging earlier investment in carbon-neutral technology. \par

Let us now discuss in detail the methodology that we follow for the construction of equilibrium. 
We start by studying the decision-making process of a single firm that is subject to the carbon pricing system and faces an irreversible investment opportunity of real-option type. We assume that the firm's technology shock process evolves according to a general It\^{o}-diffusion (cf. \cite{Miao2005, Hopenhayn1992, Luttmer2007}). Since their production induced emissions impose a negative externality on the society, a emission regulation is imposed that charges a cost -- the carbon price -- for each unit of pollutant emitted, thus decreasing firm's profits. 
As a response to this regulation, firms can potentially react by production adjustments, payment of the carbon price or abatement.
We assume that firms' abatement strategies, with the aim of lowering emissions, takes the shape of a real option problem of irreversible investment. Investment involves a sunk cost and results in the firm reducing their carbon emissions to zero. Hence, the firm can either continue paying the carbon price to account for their emissions, adjust its production or decide to exercise the option. In accordance with the existing literature (see \cite{Aid2023,Flora2020, Huang2021}), such investment should be understood as a switch to a different underlying technology, and we account for this by allowing the technology shock process to follow different dynamics after the investment.  \\
The first part of the paper thus addresses an optimal stopping problem (real option problem) that is solved using the guess-and-verify approach. We employ techniques presented in \citet{Alvarez2001}, that are tailored for dealing with general diffusions, in order to determine the firm's optimal investment time in a carbon neutral technology, thereby becoming independent of the imposed carbon price. Our general framework, in which we do not fix any explicit profit functions or particular underlying diffusions, 
is enriched by the including the characteristic of the firm leaving market due to low levels of technology (absorption in zero) or non-observable reasons (Poisson death). It is important to notice that this general formulation of the problem allows the implementation of features and parameters beyond those specified in Section \ref{ECP:Section:ExplicitModel}. This enables the study of different instances of the problem, thus drawing the focus to other parameters not captured in our specific model.\\
Under reasonable assumptions on the involved diffusions and functions, we are able to derive a complete characterization of the value function as well as the optimal investment time of the firm. The latter is given by the first time the technology shock process exceeds a threshold value $b$, reflecting the intuition that technologically advanced firms are those first to invest in carbon neutral technologies. 
Moreover, we show that the investment threshold is decreasing in the carbon price on the market: If a polluting firm faces an increased cost on their emissions, the cost-saving effect of installing an emission reduction technology becomes larger and leads to an incentive to become carbon neutral at an earlier stage (cf.~\cite{Huang2021}). \\
The single-firm optimal irreversible investment problem relates to the literature of optimal timing decisions in environmental economics, with the seminal papers of \citet{Pindyck2000, Pindyck2002} studying irreversible policy adoptions to reduce emissions of a pollutant. Since then, the real option approach to environmental investments has received much attention (cf.~\cite{Boomsma2012, Detemple2020Green, Falbo2021}). Let us mention \citet{Abadie2008, Brauneis2013, Flora2020, Insley2003} that study emission reduction investments in the presence of diffusive carbon prices, as well as \citet{Basei2023, Flora2023} as recent contributions incorporating the feature of Bayesian learning. Closest to our formulation is the work of \citet{Huang2021}, studying the problem of a company that chooses to invest in an emission reduction technology in the presence of a carbon tax. Here, the emission rate is assumed to be diffusive while the investment cost admits jumps, and the authors succeed in determining the optimal investment time to reduce its emissions. Nevertheless, it is crucial to notice that in all the aforementioned contributions of irreversible investment the carbon price is either nonexistent or exogenously given by an uncontrolled diffusion or a constant tax.  \par 
In this paper, we aim to fill this theoretical gap. To this end, we move to an aggregate level and consider a continuum of firms that are subject to idiosyncratic shocks and solve the optimal investment problem laid out in the first part. We seek to derive a stationary equilibrium consisting of an equilibrium carbon price and a stationary distribution of incumbent, polluting firms.  
First, since all firms will eventually exit (due to Poisson death or their technology shock process falling below zero) or invest (to become carbon neutral), we introduce new firms to the market via the so-called \textit{entry-condition.} Here, we equate the cost of entering and the expected benefit of entering. As usual for stationary equilibrium models, this condition guarantees the balance of the inflow and outflow of firms, such that the mass of incumbent firms is constant. \\
Second, we introduce an \textit{equilibrium condition} that equates net emissions, resulting from firms' production activities, and a constant emission level. 
Formulated in the spirit of market-clearing condition, 
the latter (fixed) parameter may be interpreted as an emission target or emission limit that is either imposed by a regulator or collectively decided on by firms (see also the discussion in Remark \ref{ECP:Remark:EquilibriumCondition}).  We highlight that a similar condition is specified in the recent \citet{Anderson2023}, studying a general equilibrium model of a government setting either quotas or taxes on emissions, and then refraining from further actions. In the case of quotas -- and in the same spirit as in this paper --  the equilibrium carbon price is determined by equating quota and total net emissions. 
The resulting carbon price should thus not be understood as a flat tax rate on emissions that is exogenously imposed by a legislative body. Instead, it is endogenously determined by the competitive actions of firms resulting from a fixed cumulative amount of emissions among them. Hence, it reflects the cost attached to carbon emissions that is necessary to achieve a given emission target and to remain in a steady state. 

It is interesting to notice that, although each firm is subject to considerable change due to its idiosyncratic noise and strategic investment decisions, the resulting equilibrium values are constant. In this regard, our work closely relates to competitive equilibrium theory, that originated in \citet{Lucas1971}, while dynamic models with entry and exit were introduced by \citet{Brock1972} and \citet{Smith1974}. While these models did not contain any firm specific stochastic elements, \citet{Jovanovic1982} first introduced a model including idiosyncratic productivity shocks. The notion of stationary equilibrium with entry and exit was then developed in the seminal papers of \citet{Hopenhayn1992} and \citet{HopenhaynRogerson1993}. Their approach serves as a tool to analyse long run behaviour of dynamic industries, resulting in equilibria with constant aggregate values. Since then, their techniques received much attention, resulting in numerous contributions,  including \citet{DixitPindyck1994}, Chapter 8, as well as \citet{Miao2005, Luttmer2007}, who study firms' investment choices and entry and exit behaviour in related frameworks. \par 
Even though the literature on optimal carbon price mechanisms is extensive (see, for example, \cite{Golosov2014, Carmona2009, Acemoglu2012, Acemoglu2016} ), our approach that combines the features of stationary equilibria, endogenous carbon pricing as well as abatement strategies presents, to our knowledge, a novelty. \\
Our modeling of abatement as a real option problem is motivated by the irreversibility in these choices. As highlighted by \citet{Abadie2008, Brauneis2013, Chesney2012}, among others, 
investments in pollution reduction are usually expensive, durable and irrevocable. Consequently, abatement strategies in order to reduce the emissions
are not decisions that firms are able to revise at a continuous rate, with typical examples including the switch to a different energy source or a different underlying technology. Notice that we are able to capture the effect of the latter policy by allowing for different dynamics of the underlying technology shock process after the investment of the firm. 
Our main contribution is then the study of the resulting stationary equilibrium and the derivation of an equilibrium carbon price. We are thus able to diverge from the assumption of an exogenously given carbon price, and characterize it as the cost that is consistent with a predetermined emission limit. As a result, we are able to examine the interplay between the irreversible investment decisions and the equilibrium carbon price.

We proceed as follows. In Section \ref{ECP:Section:OneFirmModel} we introduce the model and set the notion of equilibria. Section \ref{Section:InvestmentProblem} analyses the single-firm model and characterizes the optimal investment rule. In Section \ref{ECP:Section:Equilibrium} we aggregate and derive the existence and uniqueness of a competitive equilibrium with entry and exit.  Finally, in Section \ref{ECP:Section:ExplicitModel}, we consider a case study in which we specify the profit functions of firms and assume the technology shocks as drifted Brownian motions. Moreover, we carry out a comparative statics analysis on the explicit model, studying some of the effects the model's parameters have on the equilibrium values. In Section \ref{ECP:Subsection:RegulatorWelfare} we finally discuss the introduction of a welfare maximizing regulator, that is able to strategically set a quota on emissions.

\section{The Model}\label{ECP:Section:OneFirmModel}
This section presents our model, which is a stationary equilibrium setup.
We consider an economy consisting of a large number of firms that make entry, production and investment decisions. The output of firms is produced by either clean or dirty technology, and dirty technology contributes to greenhouse gas emissions, for example $\text{CO}_2$. Firms are subject to pollution regulation and are obliged to pay a carbon price, denoted by $c_p$, for every unit of pollutant they emit. We assume that cumulative emissions are constant over time and denoted by $E_{\max}$. The latter may be interpreted as a regulatory constraint imposed by an authority, which firms have to comply with. We comment on this particular modelling assumption in Remark \ref{ECP:Remark:EquilibriumCondition}. \\ The goal is to study a stationary equilibrium, in which all equilibrium parameters remain constant. We derive an endogenous carbon price and a stationary distribution of polluting firms. \\
Our model is posed in continuous time over an infinite time horizon, and all stochastic processes are defined on a filtered probability space $(\Omega, \mathcal{F}, \mathbb{P})$. Here, $\mathbb{F} = (\mathcal{F}_t)_{t\geq 0}$ denotes the filtration generated by a standard one-dimensional Brownian motion $(W_t)_{t \geq 0}$, satisfying the usual conditions. In the following, we set our model and notion of equilibrium. \par  

\subsection*{Firms} There exists a continuum of firms. Each firm is subject to idiosyncratic technology shocks (see for example \cite{Hopenhayn1992, Miao2005, Luttmer2007} and \cite{DixitPindyck1994} for a general discussion on idiosyncratic and aggregate shocks), which evolve according to an It\^{o}-diffusion 
\begin{align}\label{ECP:SDE:brown}
    d Z_t = \mu_1 (Z_t) dt + \sigma_1 (Z_t) d W_t,  \qquad Z_0 = z,
\end{align}
for some Borel-measurable functions $\mu_1, \sigma_1$ guaranteeing the existence of a unique strong solution to \eqref{ECP:SDE:brown} (see Assumption \ref{ECP:Assumption:CoefficientsSDE}). Firms' production depends on their current level of technology and leads to carbon emissions, that are subject to a carbon pricing system. We assume that the emissions of the single-firm can be summarized by a function $e(z; E_{\max}, c_p)$, while their profits are denoted by $\pi_1 (z; E_{\max}, c_p)$. In both cases, we stress the dependency on the overall emissions within the economy as well as the imposed carbon price. Here, we would like to highlight that one should expect a term similar to $- c_p e(z; E_{\max}, c_p)$ within the profit function $\pi_1$ of a polluting firm. Although not needed in the analysis of our general model, it reflects the idea that polluting firms are obliged to pay a fixed cost that is proportional to their emissions. Naturally, in our analysis we assume that an increase in the carbon price adversely affects firms' profits.  Moreover, we stress that each polluting firm, even though it contributes to the cumulative amount of pollution in the atmosphere, neglects its own impact. Hence, we assume that each firm, being a small player, acts as if its actions have no capability of affecting the overall level of pollution. A comparable concept is the one of a \textit{price-taking agent}, that assumes that their actions have no impact on the market price (see \cite{Maksimovic1991,Miao2005}).

\subsection*{Green Technology Adoption} Since firms suffer a cost on their emissions, they have the incentive of abatement. We assume that firms have the option to undergo an irreversible investment, that makes their production carbon neutral.
Hence, the firm can choose between paying the cost of polluting, or to install an abatement unit, that could take the form of alternative energy mix, scrubbing emissions or alternative technologies. Similar modelling assumption was made in \citet{Insley2003}, who consider the investment strategy of an electric company in a scrubber, which thus eliminates the need to purchase emission permits. Also, \citet{Flora2020} model the problem of a firm that replaces its polluting plant with a renewable one with no carbon emissions. 

The sunk cost associated to this, denoted by $c(\cdot) > 0$, could depend on the current level of technology of the firm (see, for example, \cite{Huang2021, Fan2023}) or include an installation subsidy by a potential regulator (cf.~\cite{Acemoglu2012}). Let us denote the time, at which the firm chooses to invest into becoming carbon neutral, by an $\mathbb{F}$-stopping time $\tau \geq 0$. Since a firm's investment could influence their underlying technology shock process, we assume that the latter follows a (potentially different) It\^{o}-diffusion
\begin{align}\label{ECP:SDE:green}
    d \overline{Z}_{t + \tau} = \mu_2 (\overline{Z}_{t + \tau})dt + \sigma_2 (\overline{Z}_{t + \tau}) dW_t, \quad t \geq 0, \qquad \overline{Z}_\tau = Z_\tau,
\end{align}
where $\mu_2, \sigma_2$ denote Borel-measurable functions that guarantee the existence of a unique strong solution to \eqref{ECP:SDE:green} (see Assumption \ref{ECP:Assumption:CoefficientsSDE}). To account for the dependency of the processes \eqref{ECP:SDE:brown} and \eqref{ECP:SDE:green} on their initial levels, we write $Z^z$
as well as $\overline{Z}^z$ where appropriate. The profit function of a carbon neutral firms, that is independent of the installed carbon pricing system, is denoted by $\pi_2(z; E_{\max})$. 

\subsection*{Firm Exit} We assume that firms are forced to exit the market whenever their technology shock process falls below a given level, which we, without loss of generality, assume to be given by zero (see \cite{Hopenhayn1992, Luttmer2007}). Hence, we denote the exit time by
\begin{align}\label{ECP:Def:Gamma12}
    \gamma_1 := \inf \{ t \geq 0:~Z_t = 0 \}, \qquad 
    \gamma_2 := \inf \{ t \geq \tau:~\overline{Z}_t = 0 \}.
\end{align}
Clearly, if $0$ cannot be reached by the diffusion $Z$ or $\overline{Z}$ in finite time, no absorption takes place and, as usual, we let $\inf \emptyset = + \infty$. Additionally, we assume that firms independently suffer exogenous death under the Poisson process with parameter $\eta >0$. Although not needed in order to establish the equilibrium in our model -- in contrast to related contributions (see, for example, \citet{Miao2005}) -- including this element can help account for the possibility of firms exiting the market due to factors not captured in the model.

\subsection*{The Optimal Irreversible Investment Problem} 
Firms are profit maximizers that are constrained by the pollution regulation set above. 
The firm would like to determine the optimal exercise time at which to install the abatement unit (cf.~\cite{Insley2003}). Clearly, scrubbers or alternative technologies are irrevocable commitments, while the payment of the carbon price can be utilized as need in response to the changing state variable of the firm. Since we assume that the firm has the flexibility to defer the installation until some chosen future date, this investment opportunity itself has a value. The firm then chooses the investment date in such a way to maximize its expected profits.  In summary, the value function of the single firm takes the form
\begin{align}\label{ECP:Def:ValueFunction}
    v(z) := \sup_{\tau \in \mathcal{T}} J (z,\tau),
\end{align}
where the maximization is over the set $\mathcal{T}$ of all $\mathbb{F}$-stopping times, $\mathbb{E}$ denotes the expectation with respect to the measure $\mathbb{P}$, and the objective $\mathcal{J}$ is given by
\begin{align}\label{ECP:Def:Objective1}
    \mathcal{J}(z,\tau) := \mathbb{E}
\bigg[ \int_0^{\gamma_1 \wedge \tau } e^{-(r+ \eta) t} \pi_1 (Z_t^z) dt + \Big( \int_\tau^{\gamma_2} e^{-(r+ \eta) t} \pi_2 (\overline{Z}_t^{Z_\tau^z}) dt - e^{-(r +\eta) t} c(Z_\tau^z) \Big) \one_{\{\tau < \gamma_1 \}} \bigg], 
\end{align}
where we assume that firms discount their profits with a factor $r>0$ and the term $e^{-\eta t}$ accounts for the possibility of Poisson death. We begin by rewriting the value function and, for that purpose, let 
\begin{align}\label{ECP:Def:Phi1Phi2}
    \Phi_1 (z) := \mathbb{E} \Big[ \int_0^{\gamma_1} e^{-(r + \eta)t} \pi_1(Z_t^z) dt \Big], \qquad \text{and} \qquad 
    \Phi_2 (z) := \mathbb{E} \Big[ \int_0^{\gamma_2} e^{-(r + \eta) t} \pi_2(\overline{Z}_t^z) dt \Big],
\end{align}
and are able to rewrite the objective \eqref{ECP:Def:Objective1} such that
\begin{align}\label{ECP:Def:Objective}
    \mathcal{J}(z,\tau)
    %\mathbb{E}_z \bigg[ \int_0^{\gamma \wedge \tau } e^{-r t} \pi_1 (Z_t) dt + \Big( \int_\tau^\gamma e^{-r t} \pi_2 (z_t) dt - e^{-r t} c(z_\tau) \Big) \one_{\{\tau < \gamma \}} \bigg] \\
% &= \Phi_1 (z) + \mathbb{E} \Big[ e^{-(r+\eta) \tau} \Big( \Phi_2(Z_\tau)- \Phi_1(Z_\tau) - c(Z_\tau) \Big) \one_{\{\tau<\gamma\}} \Big],
&= \mathbb{E} \Big[ \int_0^{\tau \wedge \gamma_1} e^{-(r+\eta) t} \pi_1 (Z_t^z) dt + e^{-(r + \eta) \tau} \Big( \Phi_2(Z_\tau^z) - c(Z_\tau^z) \Big) \one_{\{\tau<\gamma_1\}} \Big],
\end{align}
where we used a simple application of the strong Markov and tower property. 
We observe that the second term in the expectation on the right hand side of \eqref{ECP:Def:Objective} can be interpreted as a ``real option" of the polluting firm. If the investment is performed before the potential absorption time $\tau_1$,  the firm obtains -- after servicing the sunk cost -- the stream of revenues $\pi_2$ until it eventually leaves the market due to an inefficient technology or Poisson death. 
As we did for the profit functions $\pi_1, \pi_2$, we denote $v(z; E_{\max},c_p) = v(z; c_p) = v(z)$ where appropriate to stress the dependency of the value function on the parameters $E_{\max}$ and $c_p$.

\subsection*{New Entrants and the Entry Cost} \label{ECP:Subsection:EnteringFirms}

 Our notion of equilibrium will involve a stationary distribution of incumbent, polluting firms, which are yet to either make the irreversible investment decision or to leave the market due to their technology shock process falling below zero. Clearly, since all polluting firms will eventually leave the market or become carbon neutral, the derivation of a stationary distribution requires the existence of new entrants that strategically enter the market and take the spot of exiting firms. 

 We assume that there exists a continuum of potential new entrants, that -- upon entry -- incur a fixed sunk cost $c_e>0$ that will be specified later. After entry, the firm's initial level of technology is drawn from the distribution $\xi$ that is supported on an interval $[\underline{z}, \overline{z}]$ with $0 < \underline{z} < \overline{z}$. Notice that the latter assumption guarantees that firms do not immediately exit the market after entering it.

We notice that, while firms are not directly exiting the market due to their initial technology being below zero, it is however possible that $\overline{z} > b$. The latter implies that firms are potentially entering the market as a polluting firm and immediately choose to invest in order to become carbon neutral. 

In a competitive equilibrium, the expected benefit of entry must be equal to the entry cost; that is 
\begin{align}\label{ECP:Equ:EntryCondition}
\int_{\underline{z}}^{\overline{z}} v(z;c_p) \xi (dz) = c_e.
\end{align}
Condition \eqref{ECP:Equ:EntryCondition}, the so-called entry condition \citep[see][]{Miao2005, Hopenhayn1992,HopenhaynRogerson1993}, will be crucial when determining the equilibrium carbon price $c_p>0$, that polluting firms are obliged to pay on their emissions. More precisely, its purpose is to balance the inflow and outflow of firms, which must be equal in a stationary equilibrium to keep the mass of incumbent firms constant.

\subsection*{The Notion of Equilibria}\label{ECP:Subsection:NotionEquilibrium}
In a long run steady state, there is a stationary distribution of polluting firms $\nu$ and a constant entry rate $N$ of firms that enter the market via the mechanism we introduced in Section \ref{ECP:Subsection:EnteringFirms}. As usual in general equilibrium models, we will observe that all equilibrium variables are constant over time, even though individual firm are subject to considerable change over time. Depending on their technology shock process, polluting firms are constantly expanding, contracting, investing and even exiting, while other firms are entering. In the stationary equilibrium, however, their distribution remains constant over time. The underlying intuition lies in the fact that even though firms are subject to idiosyncratic shocks, these individual uncertainties aggregate into a stationary certainty \citep[cf.][Chapter 8]{DixitPindyck1994}. Hence, using the equilibrium distribution of active polluting firms, we are able to compute aggregate values. When considering a more specific model, this could include variables like overall output supply, capital demand, but also the cumulative emissions of the firms (see also our case study in Section \ref{ECP:Section:ExplicitModel}). The latter plays a crucial role in our analysis, as it is used to derive the equilibrium condition in our model and results in an equilibrium carbon price precisely emerging from the quantity restriction imposed on overall emissions. \\
To be more precise, we impose that in a stationary equilibrium the overall emissions $E_{max}>0$, that firms take as given in their profit functions $\pi_1$ and $\pi_2$, indeed equal the cumulative emissions of the emitting firms. Hence, the condition 
\begin{align}\label{ECP:Equ:EmissionTarget}
    E_{max} = \int_0^{b(c_p, E_{\max})} e (z;c_p,E_{max}) \nu (dz), 
\end{align}
must be satisfied, where we noticed that, due to our results from Section \ref{ECP:Section:OneFirmModel}, the support of $\nu$ is given by $(0,b(c_p, E_{\max}))$ and we highlight the dependency of the stopping threshold $b$ on the parameters $c_p$ as well as $E_{\max}$. We state the following definition, summarizing the notion of a stationary equilibrium in our model.
\begin{definition}\label{ECP:Definition:StationaryEqui}
    A stationary equilibrium consists of a carbon price $c_p^*$, an exit threshold $b^* = b (c_p^*) >0$, an entry rate $N^*$, and a distribution $\nu^*$ such that (i) $\tau_{b^*} = \inf \{ t \geq b^*: Z_t^z \geq b^*\}$ is the optimal solution to the single firm's problem \eqref{ECP:Def:ValueFunction}, (ii) the entry condition \eqref{ECP:Equ:EntryCondition} holds, (iii) the equilibrium condition \eqref{ECP:Equ:EmissionTarget} is satisfied and (iv) $\nu^*$ is an invariant distribution over $(0,b^*)$. 
\end{definition}
\begin{remark}\label{ECP:Remark:EquilibriumCondition}
    We examine the equilibrium condition described in \eqref{ECP:Equ:EmissionTarget}. It is interesting to note that this condition can be interpreted in two distinct ways. \\
    Firstly, it can be viewed as a ``consistency condition." In this interpretation, it signifies that firms' expectations regarding overall emissions remain valid and accurate within the equilibrium context. More precisely, the emissions $E_{\max}$, that firms take as given in their maximization criterion \eqref{ECP:Def:Objective}, coincide with the true overall emissions that result from their cumulative production. \\
    Alternatively, the parameter $E_{\max}$ can be seen as a regulatory constraint imposed by external authorities. This dual interpretation becomes particularly intriguing when considering models where firms do not factor overall emissions into their optimization processes. In such cases, \eqref{ECP:Equ:EmissionTarget} takes on the role of a ``compliance condition." Here, a regulatory body establishes an upper limit for the cumulative emissions of polluting firms to ensure compliance with environmental standards, and the carbon price is determined endogenously among firms. We highlight that a similar condition was specified in the recent \citet{Anderson2023}.  In their ``quota equilibrium", a regulator refrains from further actions after setting an emission limit and the price of the quota (the carbon price) is determined by equating net pollution and the pre-specified level. 
    In the subsequent section, we will demonstrate that the equilibrium entry rate is positively correlated with the maximum allowable emissions $E_{\max}$. Consequently, a regulator imposing limits on emission effectively deters potential market entrants, as their production activities could breach the stipulated emission target. \\
    At this point in our analysis, it is important to note that deriving qualitative insights regarding a potential regulator's decision criteria is challenging without further specifying the general model discussed in Section \ref{ECP:Section:OneFirmModel}. In Section \ref{ECP:Subsection:RegulatorWelfare}, we discuss the implementation of a welfare maximizing decision criterion within a more explicit model.  
\end{remark}

\section{The single-firm investment problem}
\label{Section:InvestmentProblem}
Problem \eqref{ECP:Def:ValueFunction} takes the shape of an optimal stopping problem in the field of real options theory. Dating back to the contributions of \citet{Myers1977} and \citet{Mcdonald1986}, the real options approach to irreversible investment decisions has received much attention in various problems arising in economics and finance. Here, we mention \citet{Dixit1989} as well as \citet{Pindyck1986, Pindyck1990}. 
In the case, where the underlying economic
shock process is one-dimensional -- as in our case -- explicit solutions are often feasible \citep[cf.][]{DixitPindyck1994}.  We build on their analysis and use the connection of optimal stopping and free boundary problems \citep[cf.][]{PeskirShiryaev2006} in order to identify the optimal investment rule of the single firm. \\
In the following, we denote $G(z) := \Phi_2 (z) - \Phi_1 (z) - c(z)$ and by $\mathcal{L} = \frac{1}{2} \sigma_1^2 (z) \partial_{zz} + \mu_1 (z) \partial_z$ the infinitesimal generator associated to the diffusion $Z$ of \eqref{ECP:SDE:brown}. We state the following assumption. 
\begin{assumption}\label{ECP:Assumption:(L-r)G}
   There exists a unique point $\tilde{z} \in (0, \infty)$, such that 
    \begin{align}\label{ECP:Equ:(L-r)GAssumption}
        \big( \mathcal{L} - (r+\eta) \big) G(z) = \pi_1(z) + \big( \mathcal{L} -(r+\eta)\big)[\Phi_2(z)-c(z)] \begin{cases}
            > 0, & 0 \leq z < \tilde{z}, \\
            = 0, & z= \tilde{z}, \\
            < 0, & z > \tilde{z}.
        \end{cases}
    \end{align}
\end{assumption}
Assumption \ref{ECP:Assumption:(L-r)G} is a well known criterion that is typical in optimal stopping problems in order to derive the existence and uniqueness of a point $b \in \mathbb{R}_+$ triggering the optimal stopping time (see, e.g., \citet{Alvarez2001, Falbo2021}). It is crucial to notice that, at this point, no qualitative statements are possible regarding sufficient conditions on the parameters $\mu_i, \sigma_i$ that imply the suggested shape of the function \eqref{ECP:Equ:(L-r)GAssumption} and thus of the monotonicity of $A(\cdot)$, as derived in \eqref{ECP:Equ:DerA(x)}. We remark that, under the assumption that $\mu_1 = \mu_2$, $\sigma_1 = \sigma_2$ and a constant investment cost $c(z) = I$, the condition \eqref{ECP:Equ:(L-r)GAssumption} simplifies to the  assumption $\pi_1 (z) - \pi_2(z) + (r- \eta) I < 0$ for $z > \tilde{z}$ and $\pi_1 (z) - \pi_2(z) + (r- \eta) I \geq 0$ on $z \leq \tilde{z}$. Furthermore, we mention that if Assumption \ref{ECP:Assumption:(L-r)G} is not satisfied, it can be shown that firms do not invest into a carbon neutral technology. In the case in which $0$ is natural for the diffusion $Z$ (and firms do not exit the market when their technology level is low), our analysis may lead to a stationary equilibrium in which neither entry nor exit takes place. Since this is not the scope of this work, we refrain from studying this case and stick to Assumption \ref{ECP:Assumption:(L-r)G}. \\
The following theorem summarizes our findings in the single-firm problem. Its proof can be found in Appendix \ref{ECP:Appendix:VerificationTheorem}. 
\begingroup
\allowdisplaybreaks
\begin{theorem}\label{ECP:Theorem:VerThm}
    (i) The value function $v$ of \eqref{ECP:Def:ValueFunction} takes the form
    \begin{align}
         v(z) := \begin{cases}
        0, & z \leq 0, \\
       \Phi_1(z) + \big(\Phi_2 (b) - \Phi_1 (b) - c(b)\big) \frac{\psi(z,0)}{\psi(b,0)} , & 0<z<b, \\
        \Phi_2(z) - c(z), & z \geq b,
    \end{cases}
    \end{align}
    where $b = b(c_p) \in (0, \infty)$ denotes the investment threshold triggering the optimal stopping time $\tau_b = \inf \{ t \geq 0: Z_t^z \geq b \}$. It is given by the unique solution to the nonlinear algebraic equation $A(b) = 0$, with $A(\cdot)$ given by 
    \begin{align}\label{ECP:Def:A(z)}
    A (z) := \frac{G'(z) [\psi(0) \varphi(z) - \varphi (0) \psi(z)] - G(z)[ \psi(0)\varphi'(z) - \varphi(0) \psi'(z)]}{S'(z)}.
    \end{align}
    Here, $S'$ denotes the scale density of the diffusion $Z$ and $\psi, \varphi$ denote the increasing and decreasing, respectively, fundamental solutions to the ordinary differential equation $(\mathcal{L} - (r + \eta))u = 0$. \\
    (ii) The investment threshold $b$ as well as the value function $v$ are (strictly) decreasing in the carbon price $c_p$. Moreover, the limit $b_\infty := \lim_{c_p \to \infty } b(c_p)$ exists finite. 
\end{theorem}
\endgroup

%\newpage 

\section{A Continuum of Firms and Market Equilibrium} \label{ECP:Section:Equilibrium}

In the latter section, we discussed the irreversible investment problem of a single firm, posed as an optimal stopping problem \eqref{ECP:Def:ValueFunction}. In this section, we move to an aggregate level and consider a continuum of emitting firms, that each face idiosyncratic shocks to their technology shock process and solve the investment problem introduced in Section \ref{ECP:Section:OneFirmModel}. In our forthcoming analysis, we aim to derive the long-run stationary equilibrium of the regulated market.

Hence, in order to guarantee the existence and uniqueness of a competitive equilibrium with entry and exit and a positive carbon price $c_p>0$, we state the following assumption on the exogenous entry cost $c_e$.

\begin{assumption}\label{ECP:Assumption:EntryCost}
    We state the following assumptions.
    
    (i) The entry cost $c_e$ satisfies 
    \begin{align}\label{ECP:Equ:EntryCostZero}
        c_e \leq \int_{\underline{z}}^{\overline{z}} v(z;0,E_{max}) \xi (dz).
    \end{align}
    
    (ii) We distinguish two cases. Recall that $b_\infty = \lim_{c_p \to \infty} b(c_p) >0$ exists due to Theorem \ref{ECP:Theorem:VerThm} and that $supp\{ \xi(dz)\} = [\underline{z}, \overline{z}]$. If $b_\infty > \underline{z}$, we assume
    \begin{align}\label{ECP:Equ:EntryCostbinfty}
        c_e > \lim_{c_p \to \infty} \int_{\underline{z}}^{\overline{z}} v(z;c_p,E_{max}) \xi (dz).
    \end{align}
    Otherwise, if $b_\infty \leq \underline{z}$, we then assume
    \begin{align}\label{ECP:Equ:EntryCostCpBar}
        c_e > \int_{\underline{z}}^{\overline{z}} v(z;\overline{c}_p,E_{max}) \xi (dz).
    \end{align}
    The value $\overline{c}_p>0$ is the unique solution to $b(\overline{c}_p) = \underline{z}$ (due to Theorem \ref{ECP:Theorem:VerThm}). 
\end{assumption}
     Assumption \ref{ECP:Assumption:EntryCost} is not only necessary when deriving the existence of an equilibrium in the competitive market, but also admits a clear economic interpretation. This is due to the fact that it enforces the idea that -- in any meaningful model -- the equilibrium carbon price should be such that $c_p \in (0,\infty)$, i.e.~firms do not get a positive reward for emitting pollutants. Moreover, it guarantees that a positive number of firms are entering the market that do not immediately exercise their option to become carbon-neutral. For more details we refer to the proof of Proposition \ref{ECP:Proposition:StatEquilibrium}, here we concentrate on the intuition behind Assumptions \ref{ECP:Assumption:EntryCost} $(i)$ and $(ii)$.
     
     Notice that condition $(i)$ implies that the entry cost $c_e$ is lower than the expected profit of entering firms in the absence of a carbon price $c_p$. If not fulfilled, the entry cost would thus exceed the highest possible expected benefit. This could either lead to a negative equilibrium carbon price (which is not desirable from neither a modelling nor a regulators perspective) or (if the carbon price is assumed to be positive) discourage firms from entering the market. 

     On the other hand, condition $(ii)$ implies that the entry cost is larger than the expected benefit from entering a market with the largest carbon price that still guarantees a distribution of incumbent firms. An entry cost violating condition $(ii)$ would thus lead to an imbalance, since no carbon price $c_p \in (0,\infty)$ could prevent new entrants from flooding the market. We notice that, if $b_\infty < \underline{z}$, condition \eqref{ECP:Equ:EntryCostCpBar} is not necessarily needed when deriving the stationary distribution. However, when violated, the entry condition \eqref{ECP:Equ:EntryCondition} could lead to an equilibrium carbon price $c_p$ such that $b(c_p) < \underline{z}$. This would imply that all entering firms are immediately switching to become carbon neutral. While this could be desirable from an environmental perspective, it implies a zero mass of emitting firms in the long run steady state. We exclude this trivial case here.

%\vspace{-0.6cm}
\subsection*{Existence and Uniqueness of an Equilibrium}\label{ECP:Sec:ExistenceUniquenessEqui}
Prior to presenting our central theoretical finding, which establishes the existence and uniqueness of a stationary equilibrium in our model, we offer a concise overview of the underlying rationale and the proof's methodology. 

Our proof follows similar steps as those developed in \citet{HopenhaynRogerson1993, Miao2005, DixitPindyck1994}. First, we derive the equilibrium carbon price using the entry condition \eqref{ECP:Equ:EntryCondition}. Via Assumption \ref{ECP:Assumption:EntryCost} and the derived monotonicity of the value function $v$ with respect to $c_p$ (see Theorem \ref{ECP:Theorem:VerThm}) it is straightforward to show that there exists a unique carbon price $c_p^*$ that leads \eqref{ECP:Equ:EntryCondition} to hold with equality. The equilibrium value $c_p^*$ is then used to derive the technology level $b^* = b(c_p^*)$ at which firms choose to invest into becoming carbon neutral. \\
Next, we solve for the equilibrium distribution $\nu^*$, which is, similarly as in related contributions, not a probability measure \citep[see][]{Miao2005, Hopenhayn1992, Luttmer2007}. Instead, for a given Borel set $\mathcal{B}$ on the real line, $\nu^*(\mathcal{B})$ reflects the number of polluting firms whose technology shock process falls within the set $\mathcal{B}$. Its support is given by the interval $(0,b^*)$, since polluting firms exit when their technology either falls below zero (since they are assumed to be inefficient) or exceeds the threshold $b^*$ (since they invest to become carbon neutral).    \\
Furthermore, it is crucial to notice that we are able to scale the distribution $\nu^*$ by the entry rate $N^*$ when solving for it. More precisely, one can show \citep[see][]{HopenhaynRogerson1993, DixitPindyck1994} that the stationary distribution is linearly homogeneous in the entry rate $N^*$, such that $\nu^* = N^* f(z)$ for some function $f$ to be found. The latter function is readily found using the Kolmogorov-foward equation, which is associated to the diffusion $Z$ of \eqref{ECP:SDE:brown} and complemented to include the two features entry and Poisson deaths (see \cite{Miao2005, Luttmer2007}). It follows that we can compute $f$ via ordinary differential equations, that are, depending on the position of the exit threshold $b^*$, satisfied on particular intervals. More precisely, they are given by 
\begin{align}
    - \frac{\partial}{\partial z} \big[ 
 \mu (z) f(z) \big] + \frac{\partial^2}{\partial z^2} \Big[ \frac{\sigma^2 (z)}{2} f(z) \Big] - \eta f(z) &= 0, \qquad \text{for } z \in (0, \underline{z}) \cup (\min(\overline{z},b^*) ,b^*),\label{ECP:Equ:ODENoEntry} \\ 
  - \frac{\partial}{\partial z} \big[ 
 \mu (z) f(z) \big] + \frac{\partial^2}{\partial z^2} \Big[ \frac{\sigma^2 (z)}{2} f(z) \Big] - \eta f(z) + g(z) &= 0, \qquad \text{for } z \in (\underline{z}, \min(\overline{z}, b^*)). \label{ECP:Equ:ODEEntry}
\end{align}
with boundary conditions $f(0) = f(\min (\overline{z}, b^*)) = 0$, and where $g$ denotes the density function of the entry distribution $\xi$. Notice that, since $b^*$ is endogenously determined, it is a priori unclear whether $b^* \geq \overline{z}$ or $b^* < \overline{z}$. 
The resulting stationary distribution of firms admits qualitatively different properties in these cases (notice that a fraction of entering firms is immediately switching to become carbon neutral in the latter case) and we thus distinguish them when deriving its scaled density $f$ in the proof of Proposition \ref{ECP:Proposition:StatEquilibrium} below. \\
In the last step, we use the equilibrium condition \eqref{ECP:Equ:EmissionTarget} to solve for the entry rate $N^*$. Here, it is crucial to notice that, due to the linearity of $\nu^*$ in the entry rate $N^*$, the overall emissions as determined in the integral on the right hand side of \eqref{ECP:Equ:EmissionTarget} are also linear in $N^*$. The equilibrium entry rate $N^*$ as well as the stationary distribution $\nu^* = N^* f$ are thus derived using simple calculations. 

We now state our main result. Its proof can be found in Appendix \ref{ECP:Appendix:ProofStatEqui}.

\begin{proposition}\label{ECP:Proposition:StatEquilibrium}
    There exists a unique stationary equilibrium, that includes a unique equilibrium carbon price $c_p^*>0$, an exit threshold $b^* = b(c_p^*)$, an entry rate $N^*$ and a stationary distribution $\nu^*$ such that the entry condition \eqref{ECP:Equ:EntryCondition} and the equilibrium condition \eqref{ECP:Equ:EmissionTarget} are satisfied. Moreover, $b^*$ is the threshold that triggers the optimal stopping time $\tau_{b^*}$ in the single firm problem \eqref{ECP:Def:ValueFunction}. 
\end{proposition}
We remark that our result does not require a sufficiently large parameter $\eta$ governing Poisson death. This is due to the fact that the distribution of incumbent firms is supported on the bounded interval $(0,b^*)$.  Indeed, in models with unbounded support and a non-stationary state process, a too small value of $\eta$ could lead to an exploding number of firms with large technology levels (see, for example, \cite{Miao2005, DixitPindyck1994}).

%\newpage 

\section{A Case Study with Arithmetic Brownian Motions as Technology Shocks}\label{ECP:Section:ExplicitModel}

In this section we discuss an illustrative equilibrium model of firms switching to become carbon neutral. We begin by fixing a specific model and show that the model fulfills the assumptions we imposed throughout Sections \ref{ECP:Section:OneFirmModel} and \ref{ECP:Section:Equilibrium}. In Section \ref{ECP:Section:ComparativeStaticsAnalysis} we perform a comparative statics analysis on the equilibrium parameters, analysing the sensitivity of the values on some of the underlying parameters. Furthermore, we discuss how a regulatory body could choose to constraint the overall emissions in a strategic way. Section \ref{ECP:Subsection:RegulatorWelfare} assumes a social welfare maximizing regulator, and studies the effect on the resulting equilibrium. \\
To begin with, we assume that the dynamics of $Z$ and $\overline{Z}$, as in \eqref{ECP:SDE:brown}-\eqref{ECP:SDE:green}, are given by arithmetic Brownian motions
\begin{align*}
    dZ_t &= \mu_1 dt + \sigma_1 dW_t, \quad t \geq 0, \qquad Z_0 = z, \\
    d\overline{Z}_{t +\tau} &= \mu_2 dt + \sigma_2 d W_t, \quad t \geq \tau, \qquad \overline{Z}_\tau = Z_\tau,
\end{align*}
where $\mu_i \in \mathbb{R}$ and $\sigma_i>0$, $i=1,2$. We assume that each firm has the production function
\begin{align}
    y (z,k) = D(E_{\max}) \theta z k,
\end{align}
where $\theta > 0$ is a scale parameter and $k$ denotes the capital stock of the firm, such that we are in a classical AK-model \citep[see for example][]{Walde2011}.
Moreover, we assume that the externality, i.e.~the atmospheric changes due to climate change caused by global emissions, may have a negative effect on production of each firm. Following \citet{Nordhaus2014} and \citet{Golosov2014}, we assume that damages are multiplicative and can be summarized by the damage function given by 
 \begin{align}\label{ECP:Def:DamageFunction}
     D(E_{\max}) := \exp \big( - \rho (E_{\max} - \overline{E})).
 \end{align}
\citet{Golosov2014} assume a similar structure that first translates emissions to carbon concentration, which is then translated to damages. Here, we simplify the structural form by assuming that damages are directly triggered by emissions, and $E_{\max} - \overline{E}$ measures the distance between overall emissions and a benchmark level $\overline{E}$. The latter may be chosen in such a way that the corresponding carbon concentration is that of pre-industrial times \citep[cf.][]{Golosov2014}. Notice that $\rho>0$ results in a reduced productivity of production if the level of overall emissions is above the benchmark level $\overline{E}$.

Similar to \citet{Miao2005}, we assume a simple capital structure where firms rent capital from risk-neutral investors who discount future cash flows at a constant rate $r>0$. Moreover, we assume that capital depreciates with rate $\delta>0$, such that the rental rate for firms is given by $r+\delta$. 

As in Section \ref{ECP:Section:OneFirmModel}, we assume that firms' production leads to pollution, and firms are subject to a carbon price that has to be paid according to their emissions. For simplicity, we assume that firms' emissions are proportional to their output \citep[see for example][]{Ulph1996,Krass2013, Acemoglu2016, Farzin2000}, i.e.
\begin{align}
    e(y(z,k)) = \lambda y(z,k),
\end{align}
for some $\lambda >0$. The investment cost by the firm is assumed to be constant and equal to $I - \kappa$, where $\kappa \geq 0$ denotes a subsidy of the regulator given to firms in order to give incentives for an investment to become carbon neutral.
  We assume that firms face a constant elasticity demand function \citep[see, for example][]{Bertola1998, Krass2013}, such that the market price is given by 
\begin{align*}
    p(y) = y(z,k)^{-\epsilon}, \qquad \epsilon \in (0,1).
\end{align*}
Clearly, our model could embed situations in which consumers' willingness to pay differs between products produced with different technologies (see, for example, \cite{Krass2013}). Additionally to the restructering, the firm could also benefit from possible tax incentives the legislative body will implement in order to encourage firms to become carbon neutral. Hence, we assume that polluting and carbon neutral firms are taxed with  potentially different tax rates before and after their restructure and denote them by $\tau_1 \geq \tau_2$. We note that \citet{Moyer2014} as well as \citet{Acemoglu2016} model the ``carbon tax" as a production tax which differs by type of technology, whereas we model a carbon price alongside capital income taxes, and we refer to \citet{Barrage2020} for a discussion on this topic. All in all, the firms' profit functions before and after their investment are given by  
\begin{align*}
    \pi_1 (z,k) &:= \max_k  (1-\tau_1) \big[p(y)  y(z,k) - \delta k - c_p e(y) ] -r k,   \\
    \pi_2 (z,k) &:= \max_k (1-\tau_2) \big[p(y) y(z,k) - \delta k \big] -r k.
\end{align*}
We can optimize these functions 
 and determine optimal capital levels 
\begin{align}\label{ECP:Def:CapitalDemand}
    k_1^* = \Big( \frac{(1- \epsilon) (D(E_{max}) \theta z)^{1-\epsilon} }{\delta + c_p \lambda D(E_{\max}) \theta z + r/(1-\tau_1)} \Big)^{1/\epsilon}, \qquad
        k_2^* = \Big( \frac{(1- \epsilon) (D(E_{max}) \theta z)^{1-\epsilon} }{\delta + r/(1-\tau_2)} \Big)^{1/\epsilon},
\end{align}
which, plugged into the value function, yield
\begin{align*}
    \pi_1(z) &:= \pi_1(z,k_1^*) = (1- \tau_1) \epsilon \Big( \frac{(1-\epsilon) D (E_{max}) \theta z}{\delta + c_p \lambda D(E_{max})\theta z + r/(1-\tau_1)}\Big)^{\frac{1-\epsilon}{\epsilon}}, \allowbreak
    \\
      \pi_2(z) &:= \pi_2(z,k_2^*) = (1- \tau_2) \epsilon \Big( \frac{(1-\epsilon) D (E_{max}) \theta z}{\delta + r/(1-\tau_2)}\Big)^{\frac{1-\epsilon}{\epsilon}}, \allowbreak \\
      e(z) &:= e(y(z,k_1^*) =  \lambda \Big( \frac{(1-\epsilon) D (E_{max}) \theta z}{\delta + c_p \lambda D(E_{max})\theta z + r/(1-\tau_1)}\Big)^{\frac{1}{\epsilon}}, \allowbreak
\end{align*}
and we observe that firm's profits as well as their emissions are increasing in their current technology shock (see \cite{Hopenhayn1992}). 
Regarding the entry decision of firms, we assume that the firms' initial technology values after entry are uniformly distributed, i.e. $\xi \sim \mathcal{U}([\underline{z}, \overline{z}])$ \citep[cf.][]{Miao2005}.\\
It is straightforward to verify that the proposed model satisfies the modelling assumptions made throughout the previous sections regarding the involved functions, and we provide a proof in Appendix \ref{ECP:Appendix:ModelsatisfiesAssumption}. We remark that, even in the explicit formulation of our model, a general verification of Assumption \ref{ECP:Assumption:(L-r)G} is not possible. Clearly, this would lead to imposing specific assumptions on the involved parameters in our model, which is neither a straightforward task nor leading to any qualitative insights. 

 \subsection{Comparative Statics Analysis}\label{ECP:Section:ComparativeStaticsAnalysis}
In the following, we discuss some of the implications of our case study on the equilibrium values and, especially, their sensitivity with respect to changes in the model's parameters. To begin with, we fix the base case parameter values, which are summarized in Table \ref{ECP:Table:ParametersBaseCase}.

\begin{table}[ht]
\begin{tabularx}{\textwidth}{ X >{\centering\arraybackslash}p{0.25\textwidth} >{\raggedleft\arraybackslash}p{0.25\textwidth} }
\hline
 & Parameter & Value \\
\hline
Polluting firm's shock drift &  $\mu_1$ & $0.02$\\
Polluting firm's shock volatility &  $\sigma_1$ & $0.15$\\
Carbon Neutral firm's shock drift &  $\mu_2$ & $0.02$\\
Carbon Neutral firm's shock volatility &  $\sigma_2$ & $0.15$\\
Polluting firm's tax rate & $\tau_1$ & $0.3$ \\
Carbon Neutral firm's tax rate & $\tau_2$ & $0.3$ \\
Depreciation Rate & $\delta$ & $0.1$ \\
Riskless rate & $r$ & $0.5$ \\ 
Poisson death & $\eta$ & $0.04$ \\
Entry Cost & $c_e$ & $28.6$ \\
Entry Distribution Interval & $(\underline{z}, \overline{z})$ & $(5,30)$\\
Price Elasticity & $\epsilon$ & $0.5$ \\
Scale Parameter Output & $\theta$ & $0.3$ \\
Scale Parameter Emissions & $\lambda$ & $0.05$ \\
Scale Parameter Damage & $\rho$ & $0.02$ \\
Investment Cost & $I$ & $100$ \\
Subsidy & $\kappa$ & $0$ \\
Benchmark Emission Level & $\overline{E}$ & $100$ \\
Cumulative Emissions & $E_{\max}$ & $102$ \\
\hline
\end{tabularx}
\caption{Base Case Parameter Values}
\label{ECP:Table:ParametersBaseCase}
\end{table}

We want to emphasize that for all the parameters we have selected, both in the base case model and in the upcoming sensitivity analysis, we have ensured that the condition \eqref{ECP:Equ:(L-r)GAssumption} of Assumption \ref{ECP:Assumption:(L-r)G} is consistently met. Moreover, it is essential to note that these parameter values, although chosen to align with the estimated data, serve merely as illustrative benchmarks. \\
Most of the data has been chosen to suit those assumed in related contribution as \citet{Miao2005, Golosov2014}. To allow for a clear interpretation, the entry cost as well as the interval bounds for the entry distribution in the base case model have been calibrated such that the carbon price is $c_p^* =1$. \\
In the first part of our comparative statics analysis, we maintain cumulative emissions as a fixed parameter. Here, we let $\overline{E}=100$ and $E_{\max} = 102$, such that the model depicts a 2\% breach of the benchmark level. Later, in the subsequent part (see Section \ref{ECP:Subsection:RegulatorWelfare}), we optimally select this parameter using a welfare maximization criterion of a regulator. \\
In the following, we separately discuss the effects of a change in the underlying parameters on the equilibrium values. We focus on the effect on the carbon price $c_p^*$, the investment threshold $b^*$, as well as the overall output $Y(c_p^*, E_{\max})$ and the turnover rate $T^*$ among firms. The former can be easily computed via 
\begin{align}\label{ECP:Def:OverallOutput}
   Y(c_p^*, E_{\max}) 
   := \int_0^{b^*} y(z; c_p^*, E_{\max}) \nu^*(dz) 
   = \int_0^{b^*} D(E_{\max}) \theta z k_1^*(z, c_p^*, E_{\max}) \nu^* (dz),
\end{align}
where $k_1^*$ denotes the optimal capital demand as in \eqref{ECP:Def:CapitalDemand}. The turnover rate should be understood as the ratio between the entry rate to the mass of incumbent firms in equilibrium. We notice that the latter can be computed via 
\begin{align*}
    M^* := \int_0^{b^*} \nu^* (dz) = N^* \int_0^{b^*} f(z) dz,
\end{align*}
such that the turnover rate writes as 
\begin{align*}
     T^* = \frac{N^*}{M^*} = \frac{1}{\int_0^{b^*} f(z) dz}.
\end{align*}

It is interesting to observe that \eqref{ECP:Def:OverallOutput} displays an \textit{output effect} of pollution regulation. As a response to the imposed carbon pricing system, firms decrease their production output in order to comply with the given emission target. More stringent regulation, imposed through a lowered emission limit, thus leads to a decrease in the cumulative output of firms.

\subsection*{Sensitivity with respect to the Diffusion Coefficients} 
We investigate the impact of a change in the coefficients governing the underlying diffusions. Our findings are summarized in Table \ref{ECP:Table:CompStatDiffusionCoefficients}.\\
We note that when the technology growth parameter $\mu_1$ for polluting firms decreases, the exit threshold decreases correspondingly. This is a logical outcome since the profit function $\pi_1$ for polluting firms increases as technology levels improve. A smaller drift in this context leads to reduced expected operating profits, leading firms to invest earlier in achieving carbon neutrality. Consequently, the exit threshold $b$ is lowered. Moreover, we observe that in equilibrium an increased technology growth among polluting firms results in a rising carbon price. This can be attributed to the fact that as firms experience greater technological advancements, their output and emissions also increase. To counteract this effect, in order to keep the cumulative emissions in balance with the equilibrium condition, a higher equilibrium carbon price encourages firms to opt for lower output levels. \\
We also observe that a higher technology growth rate increases the turnover rate. It is worth noting that this increased turnover rate may have different causes: firms may exit the market due to decreased efficiency, as indicated by their technology shock process falling below zero, or they may become more efficient, leading to a higher number of firms adopting carbon-neutral technology and leaving the market while still operational. In this scenario, we find that, although the effect is minor, the latter factor seems to dominate.  \\
Regarding changes in the technology growth parameter $\mu_2$ for carbon-neutral firms, we find that this has a negligible effect on the carbon price, even though it influences the exit threshold as expected. An increase in the technology growth of carbon-neutral firms boosts the expected profit for firms following their investment. This incentive encourages firms to invest sooner, resulting in a decrease in the exit threshold. \\
Regarding a shift in the volatility of the technology shock process, we observe only minor effects. While a larger volatility of polluting firms' technology shock process increases the turnover rate and leads to delayed investment, we find no significant effect of a shift in the parameter $\sigma_2$ on the equilibrium values. \par
\begin{table}
\begin{tabularx}{\textwidth}{ X >{\centering\arraybackslash}p{0.17\textwidth} >{\centering\arraybackslash}p{0.24\textwidth} >{\centering\arraybackslash}p{0.17\textwidth} >{\raggedleft\arraybackslash}p{0.17\textwidth}}
\hline
 & Carbon Price & Investment Threshold & Turnover Rate & Overall Output \\
\hline
Base case & $1.00$ & $32.78$ & $0.0403$& $ 2040$\\[0.1cm]
$\mu_1 = 0.01$ & $0.99$ & $32.72$ & $0.0401$ & $2040$\\
$\mu_1 = 0.03$ & $1.01$ & $32.84$ & $0.0406$ & $2040$\\[0.1cm]

$\mu_2 = 0.01$ & $1.00$ & $32.93$ & $0.0403$ & $2040$\\
$\mu_2 = 0.03$ & $1.00$ & $32.62$ & $0.0403$ & $2040$\\
[0.1cm]
$\sigma_1 = 0.2$ & $1.00$ & $32.91$ & $0.0404$ & $2040$\\
$\sigma_1 = 0.25$ & $1.00$ & $33.04$ & $0.0406$ & $2040$\\[0.1cm]
$\sigma_2 = 0.2$ & $1.00$ & $32.77$ & $0.0403$ & $2040$\\
$\sigma_2 = 0.25$ & $1.00$ & $32.77$ & $0.0403$ & $2040$\\
\hline
\end{tabularx}\caption{Comparative Statics with respect to the diffusion coefficients}
\label{ECP:Table:CompStatDiffusionCoefficients}
\end{table}

\subsection*{Sensitivity with respect to the Tax Rates $\tau_1$ and $\tau_2$}\label{ECP:Sec:CompStatTaxRates}
\begin{figure}[!b]
    \centering
    \includegraphics[width=0.45\textwidth]{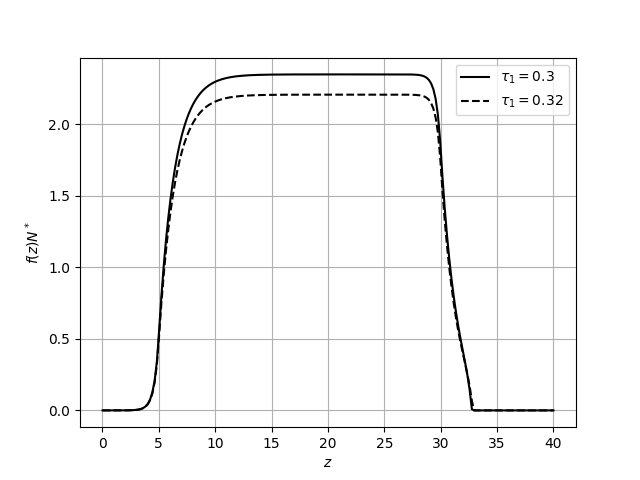} ~
    \includegraphics[width=0.45\textwidth]{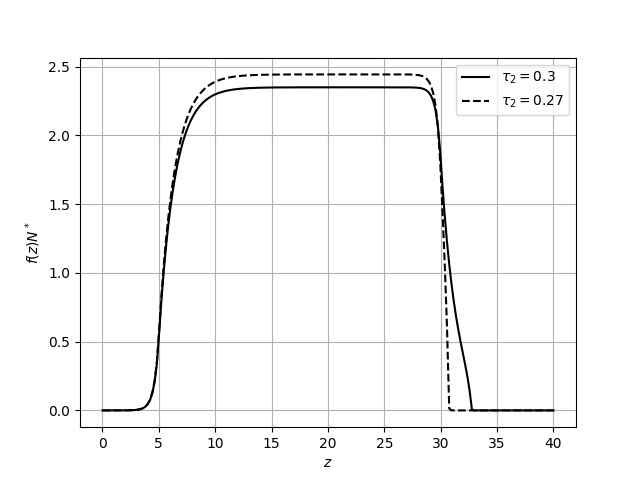}
    \caption{Equilibrium density of polluting firms with respect to a change in the tax rates $\tau_1$ and $\tau_2$, respectively.}
    \label{ECP:Fig:DensTau12}
\end{figure}

We now study the effect of a shift in the corporate tax rates of firms. A potential regulator or legislative body could install different tax rates for polluting and carbon neutral firms, in order to encourage firms to invest earlier into becoming carbon neutral. However, in contrast to direct prohibitions on the use of certain technologies, setting a certain corporate tax level (as well as the carbon price) is an indirect tool that tries to provide incentives for firms to switch to the ``right" technology. We note that this regulation policy could be installed by either increasing taxes for polluting firms, or by installing tax benefits (and thus decreasing the corporate tax) for carbon neutral firms. The observed effects on the equilibrium parameters are summarized in Table \ref{ECP:Table:CompStatTaxRates}. \\
Regarding an increase in the tax rates for polluting firms, we observe the following. Even though a higher tax rates should provide an incentive to adopt cleaner technology earlier, we observe a counter intuitive effect in our model. The reason lies in the fact that a tax increase has two separate effects. First, it decreases corporate profits and thus leads firms to invest earlier in alternative technologies. On the other hand, it decreases the expected profit that firms face when considering entering the market. The latter effect leads to less entry and a smaller number of incumbent firms, which can be observed in Figure \ref{ECP:Fig:DensTau12}, which plots the mass of polluting firms for different tax rates $\tau_1$. \\
\begin{table}[t]
\begin{tabularx}{\textwidth}{ X >{\centering\arraybackslash}p{0.17\textwidth} >{\centering\arraybackslash}p{0.24\textwidth} >{\centering\arraybackslash}p{0.17\textwidth} >{\raggedleft\arraybackslash}p{0.17\textwidth}}
\hline
 & Carbon Price & Investment Threshold & Turnover Rate & Overall Output \\
\hline
Base case & $1.00$ & $32.78$ & $0.0403$& $ 2040$\\[0.1cm]
$\tau_1 = 0.32$ & $0.94$ & $32.87$ & $0.0403$ & $ 2040$\\
$\tau_1 = 0.35$ & $0.86$ & $33.02$ & $0.0403$ & $2040 $\\[0.1cm]
$\tau_2 = 0.27$ & $1.00$ & $30.76$ & $0.0413$ & $2400 $\\
$\tau_2 = 0.25$ & $1.01$ & $29.49$ & $0.0433$ & $2400 $\\[0.1cm]
$\kappa = 10$ & $1.00 $ & $30.19$ & $0.0421$ & $2040$\\
$\kappa = 20$ & $1.05$ & $27.28$ & $0.0479$& $2040$\\
\hline
\end{tabularx}
\caption{Comparative Statics with respect to the tax rates and the investment subsidy}
\label{ECP:Table:CompStatTaxRates}
\end{table}
\noindent Since less firms are on the market, the competition among firms to achieve the given cumulative emissions decrease. Individual firms are able to produce and pollute more, at the same time the carbon price decreases, which increases firms' profits. This leads firms to stay longer in the market as a polluting firm, and the the exit threshold $b$ increases. In the current parameter constellation, this latter effect seems to outweigh the former. \\
Concerning the effect on the carbon price, we notice the following. A tax increase on polluting firms is a targeted policy that aims to decrease firms profits and thus encourage them to invest into a carbon neutral technology. This, as explained above, distorts the polluting sector by decreasing the willingness to enter and thus the mass of incumbent firms. In equilibrium, the carbon price is chosen such that a given emission target is met. If taxes increase, less firms are in competition for the cumulative emissions and it follows that a lower carbon price is needed to keep firms emissions within the stipulated emission target. This market size effect, caused by the presence of distortionary taxes, is also discussed in \citet{Barrage2020}.  \\ 
As for a decrease in the tax rate for carbon neutral firms, we note the following. Similarly as the policy discussed above, which however targets the polluting firms, a tax reduction for carbon neutral firms aims at providing incentives to invest in cleaner technology and thus accelerate the process of firms becoming carbon neutral. Equivalently, these tax incentives can be understood as production subsidies for green firms, as implemented in a number of US States (see also \cite{Drake2016}). The resulting, desired effect on firms' abatement strategy can be observed in our model, as the exit threshold indeed decreases. Furthermore, the turnover rate increases, which can be explained by the increased amount of firms switching to a carbon neutral technology, due to the tax policy making it financially untenable for highly polluting firms to continue operating without making significant changes to their practices. \\
Interestingly, the chosen parameter constellation leads to an exit threshold that is lower than the upper interval bound $\overline{z}$ up to which firms initial technology level is drawn. Hence, a lower tax rate for carbon neutral firms may lead some firms to enter the market and -- if their initial technology is sufficiently high -- immediately become carbon neutral. In this regard, a lower tax rate encourages a higher entry into the market, since firms observe that it is profitable to become carbon neutral and thus benefit from a tax cut for smaller values of the technology shock process. \\
The carbon price seems to be relatively robust with respect to a change in the tax rates $\tau_2$. Intuitively, this results from the fact that this tax policy targets carbon neutral firms, while the carbon price in equilibrium is determined endogenously among the polluting firms. The small effect observed in Table \ref{ECP:Table:CompStatTaxRates} can be explained by the increased expected profit that potential entrants face when making their entry decision, especially when it becomes profitable to enter the market and directly switching to a carbon neutral technology. A slightly higher carbon price counteracts this in order to balance the entry condition and to prevent the market from getting flooded by new market entrants.

\subsection*{Sensitivity with respect to the Investment Subsidy}
An alternative to a targeted tax increase/decrease in order to encourage firms to invest in a carbon neutral technology (as discussed previously) is by offering a subsidy to firms in order to (partially) offset their investment cost. Examples include clean investment subsidies offered by most US states, that promote investments in green technologies (see also \cite{Drake2016}). 
In their models, \citet{Acemoglu2012} and \citet{Acemoglu2016} show that an optimal environmental regulation always uses a carbon price/tax and a subsidy to clean research. While the former should control for the carbon emissions, the latter directs innovation towards the clean sector. We observe that it is methodologically related to a tax cut for carbon neutral firms, as it rewards environmentally beneficial behaviour instead of punishing the polluting firms. Consequently, the observed effects on the equilibrium values for the carbon price, exit threshold and turnover rate are related to the ones discussed above, such that the region where the clean technology is chosen is expanded (see also \cite{Krass2013}). The equilibrium density of polluting firms can be observed in Figure \ref{ECP:Fig:SubsidyPoissonDeath}. \par  

\subsection*{Sensitivity with respect to Poisson Death}
\begin{figure}[!b]
    \centering
    \includegraphics[width=0.45\textwidth]{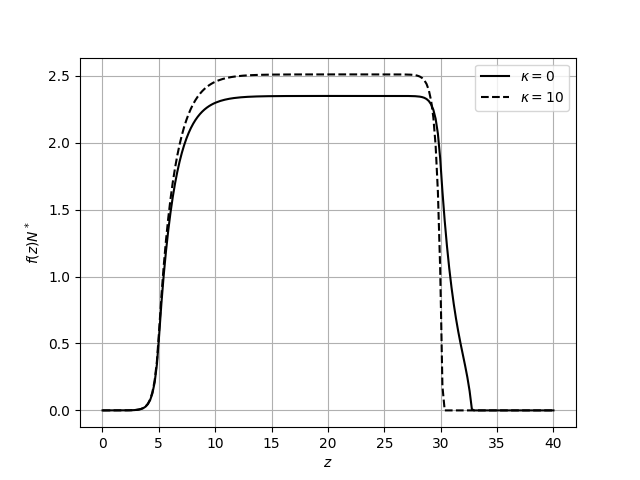} ~
    \includegraphics[width=0.45\textwidth]{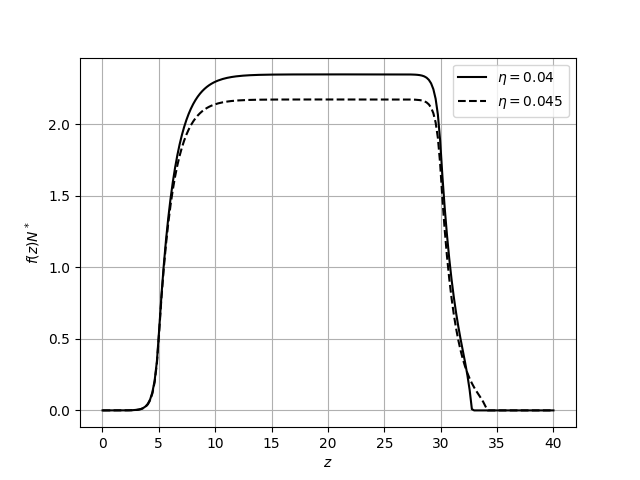}
    \caption{Equilibrium density of polluting firms with respect to a change in the subsidy $\kappa$ and in the parameter of Poisson death $\eta$, respectively.}
    \label{ECP:Fig:SubsidyPoissonDeath}
\end{figure}

According to the the size of the parameter of Poisson death, firms are randomly subject to a shock that leads them to leave the market.  
\begin{table}[t]
\begin{tabularx}{\textwidth}{ X >{\centering\arraybackslash}p{0.17\textwidth} >{\centering\arraybackslash}p{0.24\textwidth} >{\centering\arraybackslash}p{0.17\textwidth} >{\raggedleft\arraybackslash}p{0.17\textwidth}}
\hline
 & Carbon Price & Investment Threshold & Turnover Rate & Overall Output \\
\hline
Base case & $1.00$ & $32.78$ & $0.0403$& $ 2040$\\[0.1cm]
$\eta = 0.035$ & $1.06$ & $31.46 $ & $0.0359 $ & $ 2040$\\
$\eta = 0.045$ & $0.94$ & $34.11$ & $0.0451 $ & $2040$\\[0.1cm]
\hline
\end{tabularx}
\caption{Comparative Statics with respect to the parameter of Poisson death}
\label{ECP:Table:CompStatPoisson}
\end{table}
We observe that an increasing parameter $\eta$ that governs the rate at which firms randomly leave the market, leads -- as expected -- to a higher turnover rate on the market and to a lower mass of polluting firms, as observed in Figure \ref{ECP:Fig:SubsidyPoissonDeath}. This reduces competition for firms, and consequently the carbon price decreases. 

\subsection*{Sensitivity with respect to Emission-related Parameters}
Here, we discuss the sensitivity of the equilibrium parameters with respect to the emission related parameters, namely the emission target $E_{\max}$ and the scale parameters $\lambda$ and $\rho$. \\
First, we notice that -- differently from the comparative statics analysis so far -- a shift in the values $E_{\max}$ or $\lambda$ does influence the overall production in equilibrium. This is due to the fact that the firm's emissions are assumed to be proportional to their output with parameter $\lambda$, i.e.~$e(z) = \lambda y(z)$, such that the equilibrium condition \eqref{ECP:Equ:EmissionTarget} rewrites as
\begin{align*}
    \frac{E_{\max}}{\lambda} = \int_0^{b(c_p^*, E_{max})} y (z; c_p^*, E_{max}) \nu^* (dz),
\end{align*}
where the latter integral thus denotes the aggregate production of the polluting firms. Hence, even though firms produce distinct goods facing a constant elasticity of demand function, they indirectly compete on quantities as the economies output is limited through the pollution regulation. It follows that an increase in the overall permitted emissions leads to an increase in overall output, while it decreases if the production of firms leads to higher emissions (see also \cite{Anand2020}). The latter can be observed in Table \ref{ECP:Table:CompStatEmissionParameters}. 
\begin{table}[t]
\begin{tabularx}{\textwidth}{ X >{\centering\arraybackslash}p{0.17\textwidth} >{\centering\arraybackslash}p{0.24\textwidth} >{\centering\arraybackslash}p{0.17\textwidth} >{\raggedleft\arraybackslash}p{0.17\textwidth}}
\hline
 & Carbon Price & Investment Threshold & Turnover Rate & Overall Output \\
\hline
Base case & $1.00$ & $32.78$ & $0.0403$& $ 2040$\\[0.1cm]
$E_{\max} = 100$ & $1.03 $ & $31.33$ & $0.0409$ & $ 2000$\\
$E_{\max} = 105$ & $0.96 $ & $35.10 $ & $0.0401 $ & $2100$\\[0.1cm]

$\lambda = 0.049$ & $1.02$ & $32.78$ & $0.0403$ & $ 2081$\\
$\lambda = 0.051$ & $0.98 $ & $32.78$ & $0.0403$ & $ 2000$\\
[0.1cm]
$\rho = 0.01 $ & $1.01 $ & $32.04 $ & $0.0405 $ & $2040$\\
$\rho = 0.03$ & $0.99$ & $33.53 $ & $0.0402 $ & $2040$\\[0.1cm]
\hline
\end{tabularx}
\caption{Comparative Statics with respect to the emission target and the scale parameters of emissions and damage}
\label{ECP:Table:CompStatEmissionParameters}
\end{table}
\begin{figure}[!b]
    \centering
    \includegraphics[width=0.45\textwidth]{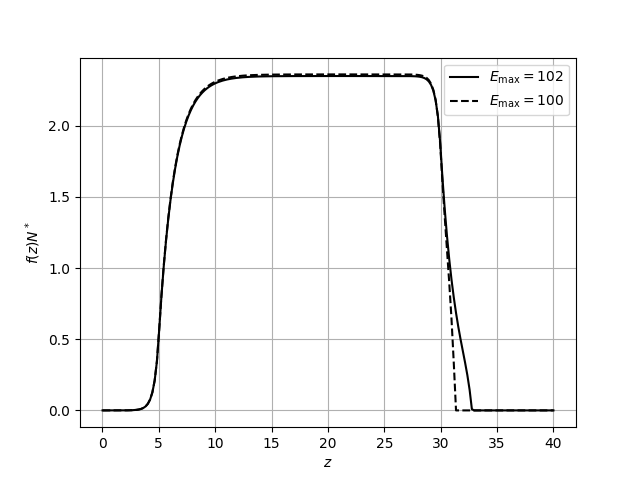} ~ 
    \includegraphics[width=0.45\textwidth]{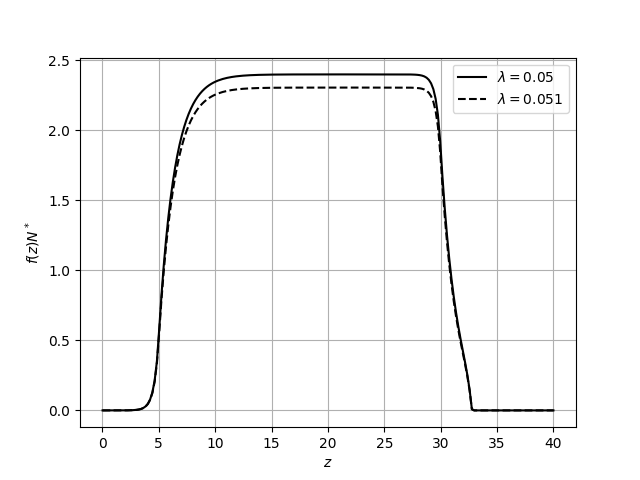} 
    \caption{Equilibrium density of polluting firms with respect to a change in the overall emissions $E_{\max}$ and the scale parameter $\lambda$, respectively.}
    \label{ECP:Fig:EmaxLambda}
\end{figure}

Regarding a shift in the overall emissions $E_{\max}$ we observe the following. Assuming a lower value of $E_{\max}$ translates to a more rigorous target for the cumulative emissions. First, as discussed above, it follows that the overall production output decreases. Furthermore, the competition for firms to emit increases, which leads to an increase in the equilibrium carbon price. Consequently, firms are encouraged to invest into becoming carbon neutral, such that the exit threshold $b$ decreases and the turnover rate increases accordingly. In Figure \ref{ECP:Fig:EmaxLambda} we observe that the shift of overall emissions mostly affects firms with high values of technology. While the mass of firms with low or intermediate values of technology barely changes, the reduction of overall emissions is mainly achieved by firms with high values of technology, as they choose to invest into a carbon neutral technology at an earlier stage. This regulation consequence of both an output reduction as well as a pollution abatement effect (through the lowered investment threshold) was also observed in other, related contributions (see, for example, \cite{Anand2020}). \\
The parameter $\lambda$ serves as a measure on how emission intensive the production of firms is -- larger values of $\lambda$ imply higher emissions per unit of output. Thus, as explained above, the equilibrium output is decreasing in $\lambda$. In Table \ref{ECP:Table:CompStatEmissionParameters} we observe that the carbon price is decreasing in the carbon intensity of production. In some sense, the carbon price acts as a counterpart to the carbon intensity, since $1.11 \times 0.045 \approx 0.91 \times 0.055 \approx 0.05$, where the latter is precisely the carbon intensity in the base case model. Hence, in order to achieve a cumulative emission target, the carbon price decreases in the carbon intensity in order to balance the cost of production of the firms. Interestingly, this leads to the same exit threshold and turnover rate as in the base case model. While the overall emissions stay constant, the decrease in output mainly results from the smaller mass of incumbent firms, as observed in Figure \ref{ECP:Fig:EmaxLambda}. \\
The parameter $\rho$, appearing in the damage function \eqref{ECP:Def:DamageFunction}, scales the effect of the damage that the breach of the benchmark level has on the production of the firms. It is crucial to notice that we assumed the damage function to appear in both the profit functions of the polluting as well as the carbon neutral firms, such that an increase in damage does not necessarily lead firms to invest earlier since they remain affected by the damage function. Here, we observe that the decrease in profit leads to less competition in the market, which leads to a lowered carbon price and an increased exit threshold. 

\subsection*{Sensitivity with respect to the Entry Distribution and the Entry Cost}
\begin{table}[t]
\begin{tabularx}{\textwidth}{ X >{\centering\arraybackslash}p{0.17\textwidth} >{\centering\arraybackslash}p{0.24\textwidth} >{\centering\arraybackslash}p{0.17\textwidth} >{\raggedleft\arraybackslash}p{0.17\textwidth}}
\hline
 & Carbon Price & Investment Threshold & Turnover Rate & Overall Output \\
\hline
Base case & $1.00$ & $32.78$ & $0.0403$& $ 2040$\\[0.1cm]
$c_e = 27$ & $1.10 $ & $32.14 $ & $0.0405 $ & $2040$\\
$c_e = 30$ & $0.92$ & $33.35 $ & $0.0402$ & $2040$\\[0.1cm]
$ \underline{z} = 3$ & $ 0.93$ & $33.28 $ & $0.0402 $ & $2040$\\
$\underline{z} = 7$ & $1.06 $ & $32.40 $ & $0.0404 $ & $2040$\\
\hline
\end{tabularx}
\caption{Comparative Statics with respect to the entry cost and entry distribution}
\label{ECP:Table:CompStatEntry}
\end{table}
\begin{figure}[b]
    \centering
    \includegraphics[width=0.45\textwidth]{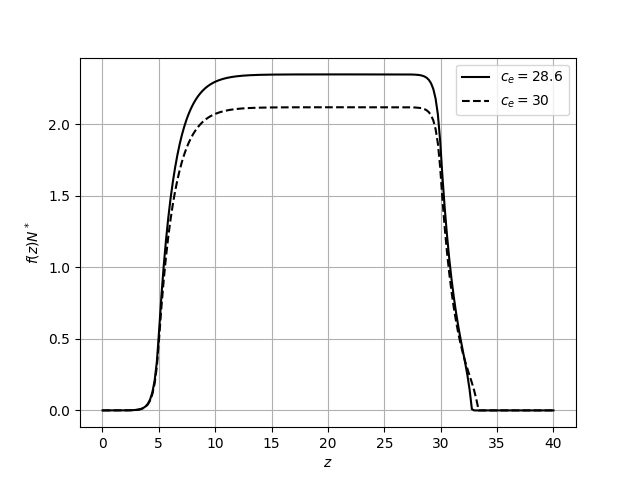} ~
    \includegraphics[width=0.45\textwidth]{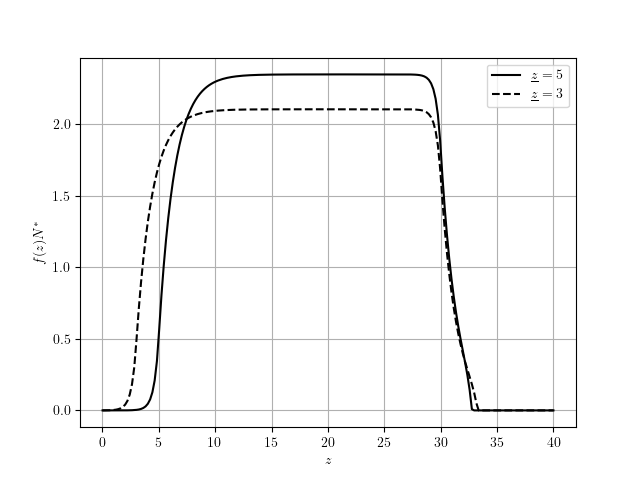}
    \caption{Equilibrium density of polluting firms with respect to a change in the entry cost and the entry dsitribution}
    \label{ECP:Fig:EntryCostDistr}
\end{figure}
Table \ref{ECP:Table:CompStatEntry} summarizes the effect on the equilibrium parameters when subject to a change in the entry cost $c_e$ as well as the lower interval bound $\underline{z}$ of the entry distribution. Notice that both parameters do not affect the running profit functions of any of the incumbent firms, both polluting and carbon neutral. However, they play a huge part in the entry decision of potential entrants. A  decreasing entry cost thus leads to a lower barrier to entry and to more firms willing to enter, as reflected in the increased turnover rate and mass of incumbent firms, as indicated in Figure \ref{ECP:Fig:EntryCostDistr} (see \cite{Miao2005, Luttmer2007}). Consequently, the enhanced competition leads to a larger carbon price, which results in  incumbent firms aiming to become carbon neutral and thus invest at an earlier stage. \\
A similar effect can be seen for the increase in the lower interval bound of the entry distribution $\xi$, that is uniform on the interval $[\underline{z}, \overline{z}]$. Similar to a decrease in the entry cost, increasing $\underline{z}$ leads to a larger expected profit for potential entrants. As seen above, this increases the turnover rate and competition on the market, which results in a larger carbon price and a lower investment threshold.

\subsection{A Welfare Maximizing Regulator}\label{ECP:Subsection:RegulatorWelfare}
Here, we come back to the concept introduced in Remark \ref{ECP:Remark:EquilibriumCondition}, which posits that the overall emissions may be determined by deliberate decisions made by legislative bodies or environmental regulatory agencies. This concept is closely related with the principles underlying a cap-and-trade market framework, wherein a regulatory authority makes strategic decisions regarding the allocation of emission allowances to the market. In the context of our model, the cumulative emissions can be viewed as a regulatory constraint imposed on the collective emissions of firms. \\
Within the literature, a common approach entails the perspective of a ``welfare-maximizing" regulator, as in \citet{Barrett2001}, \citet{Colla2012}, \citet{Ulph1996}, and \citet{Germain2004}. In this framework, the regulator, when setting a limit on overall emissions, aims to balance the societal benefits derived from emissions reduction and the economic losses incurred due to the constraints imposed. In the mentioned contributions, it is assumed that the regulator maximizes a function encompassing aggregate production, adjusted for capital consumption, and the ``cost of pollution-induced damages", i.e.
\begin{align}\label{ECP:Def:RegulatorProblem}
    \max_{E_{\max}} \Big( Y(E_{\max}) - r K (E_{\max}) -  \Gamma E_{\max}^{1+w} \Big),  
\end{align}
where $\Gamma, w >0$ weigh the cost of pollution-induced damages (and thus translate a unit of emissions into a monetary unit), $Y$ denotes the aggregate supply of firms as in \eqref{ECP:Def:OverallOutput} and 
\begin{align*}
    K(E_{\max}) &= \int_0^{b^*(c_p, E_{\max})} k_1^*(z, E_{\max} ) \nu^* (dz) = \int_0^{b^*(c_p, E_{\max})} k_1^*(z, E_{\max} ) N^* f^*(z) dz 
    \end{align*}
denotes their consumption of capital. The proof of Proposition \ref{ECP:Proposition:StatEquilibrium} reveals that the equilibrium entry rate $N^*$ is determined via
\begin{align*}
     N^* &= \frac{E_{max}}{\int_0^{b^*(c_p, E_{\max})} e (y (z;c_p,E_{max})) f^*(z) dz} = \frac{E_{max}}{ \lambda \int_0^{b^*(c_p, E_{\max})} y (z;c_p,E_{max}) f^*(z) dz},
\end{align*}
and hence, in equilibrium, we obtain
\begin{align*}
 &Y(E_{\max}) - r K (E_{\max}) - \Gamma E_{\max}^{1+w} \\
 &= \frac{E_{max} \int_0^{b^*(c_p, E_{\max})} y(z; c_p, E_{\max} ) f^*(z) dz }{ \lambda \int_0^{b^*(c_p, E_{\max})} y (z;c_p,E_{max}) f^*(z) dz} - \frac{E_{max} r \int_0^{b^*(c_p, E_{\max})} k_1^*(z; c_p, E_{\max} ) f^*(z) dz }{ \lambda \int_0^{b^*(c_p, E_{\max})} y (z;c_p,E_{max}) f^*(z) dz} - \Gamma E_{\max}^{1+w} \displaybreak[0] \\
 &= E_{\max} \Big( \frac{1}{\lambda} - \frac{r \int_0^{b^*(c_p, E_{\max})} k_1^*(z; c_p, E_{\max} ) f^*(z) dz }{ \lambda \int_0^{b^*(c_p, E_{\max})} y (z;c_p,E_{max}) f^*(z) dz} - \Gamma E_{\max}^w \Big).
\end{align*}
\begin{table}[t]
\begin{tabularx}{\textwidth}{ X  >{\centering\arraybackslash}p{0.1\textwidth}>{\centering\arraybackslash}p{0.15\textwidth} >{\centering\arraybackslash}p{0.24\textwidth} >{\centering\arraybackslash}p{0.16\textwidth} >{\raggedleft\arraybackslash}p{0.07\textwidth}}
\hline
 & Emissions & Carbon Price & Investment Threshold & Turnover Rate & Output \\
\hline
Base case & $102.0$ & $1.00$ & $32.78$ & $0.0403$& $ 2040$\\[0.1cm]
$\Gamma = 0.0985$ &$101.3$ & $1.01$ & $32.28$ & $0.0404$ & $2026$\\
$\Gamma = 0.0990$ & $100.8$ & $1.02$ & $31.91$ & $0.0406$ & $2016$\\
\hline
\end{tabularx}
\caption{Comparative Statics for the welfare maximizing equilibrium values with respect to the scale parameter $\Gamma$}
\label{ECP:Table:CompStatWelfareMaxEqui}
\end{table}
Due to the dependencies of the firm's production $y$, their capital demand $k_1^*$ (as in \eqref{ECP:Def:CapitalDemand}), and the investment threshold $b^*$ on the overall emissions $E_{\max}$, a closed form solution for the maximization problem \eqref{ECP:Def:RegulatorProblem} does not seem feasible. To this end, we focus solely on the numerical analysis here. In accordance with \citet{Pommeret2013, Pindyck2002, Anand2020}, and \citet{Colla2012}, Section 4, we set $w = 1$ so to obtain a quadratic cost of damages. Our numerical analysis suggests that this indeed guarantees the existence of a maximum in the regulator's optimization problem \eqref{ECP:Def:RegulatorProblem}, irrespective of the chosen parameter $\Gamma>0$. \\
Clearly, the goals of the regulator are to keep the output at a high level, while minimizing the capital consumption as well as economic damages. These goals, as observed earlier, stand necessarily in conflict as the cumulative output strictly decreases in the cap on overall emissions. We highlight that also other criterions than that of \eqref{ECP:Def:RegulatorProblem} are possible to be implemented within the current framework. Including firms cumulative profits, surplus from the net revenue of collecting the carbon price or a measure of consumer surplus are possible ways to extend the proposed objective (see, for example, \cite{Anand2020, Krass2013}).  \\
Note that the value of overall emissions $E_{\max}$ is now chosen endogenously as well, and thus becomes part of the equilibrium variables. We refer to the equilibrium, that involves the emission target set by the regulator via the optimality criterion \eqref{ECP:Def:RegulatorProblem}, as the \textit{welfare maximizing equilibrium}. In order to allow for a more comprehensible study of the latter, and to compare it with the equilibrium values we obtained in the previous comparative statics analysis for fixed overall emissions, we calibrate $\Gamma$ such that the benchmark welfare maximizing equilibrium is such that $E_{\max}^* = 102$, as in our base case model.  
We do not repeat the extensive comparative statics analysis studied before, but mainly focus on the emission related parameters $\lambda, \rho$ (as studied in Section \ref{ECP:Section:ComparativeStaticsAnalysis}) and the scale parameter $\Gamma$ that measures the cost of damage induced by pollution. 

\begin{figure}[t]
    \centering
    \includegraphics[width=0.45\textwidth]{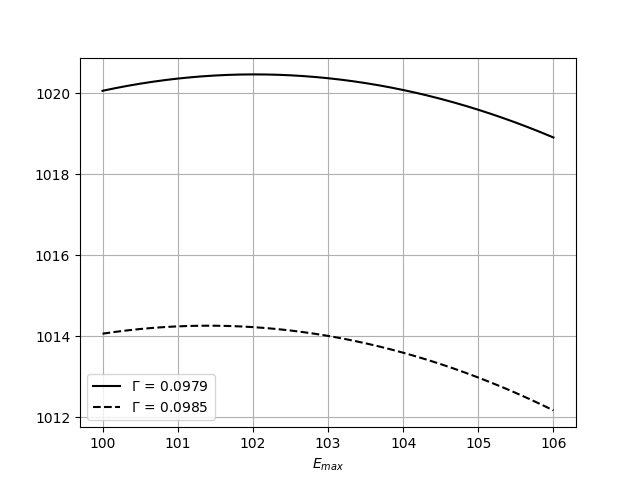}~
    \includegraphics[width=0.45\textwidth]{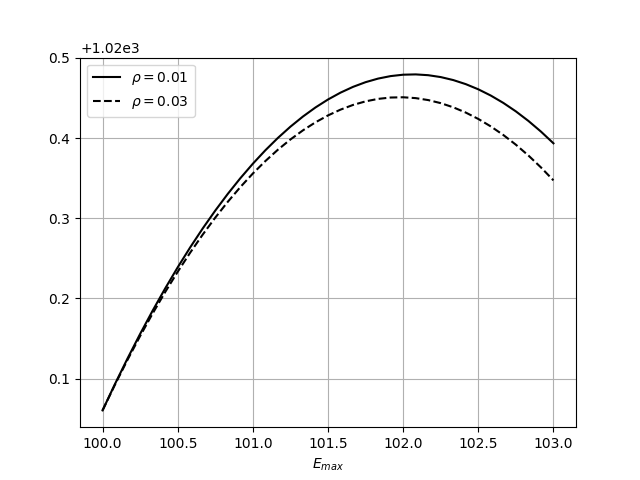}
    \caption{The social welfare for different values of the scale parameters $\Gamma$ and $\rho$, respectively.}
    \label{ECP:Fig:Regulator}
\end{figure}

Increasing the pollution damage factor clearly leads to an increased cost for the regulator. In Table \ref{ECP:Table:CompStatWelfareMaxEqui} we observe that the regulator reacts by decreasing the emission target $E_{\max}$, even though this has the consequence of a lower overall output level. As observed in Section \ref{ECP:Section:ComparativeStaticsAnalysis}, decreasing the emission target leads to an increased carbon price, a lower investment threshold and a higher turnover rate. We conclude that a larger (negative) effect of pollution should be addressed by setting a stricter emission limit on the regulated firms. Moreover, we find that the higher the marginal damage, the lower the social welfare, as observed in Figure \ref{ECP:Fig:Regulator}. This intuitive result complements the findings in \citet{Colla2012, Krass2013}. 
\begin{table}[t]
\begin{tabularx}{\textwidth}{ X  >{\centering\arraybackslash}p{0.1\textwidth}>{\centering\arraybackslash}p{0.15\textwidth} >{\centering\arraybackslash}p{0.24\textwidth} >{\centering\arraybackslash}p{0.16\textwidth} >{\raggedleft\arraybackslash}p{0.07\textwidth}}
\hline
 & Emissions & Carbon Price & Investment Threshold & Turnover Rate & Output \\
\hline
Base case & $102.0$ & $1.00$ & $32.78$ & $0.0403$&$2040$\\
\hdashline
\multirow{2}{4em}{${\lambda=0.049}$} & $102.0$& $1.02$ & $32.78$ & $0.0403$ & $ 2081$\\
& $104.1 $ & $0.99$ & $34.41$ & $0.0401$ & $2125$ \\
\hdashline
\multirow{2}{4em}{${\lambda=0.051}$} & $102.0$ & $0.98 $ & $32.78$ & $0.0403$ & $ 2000$\\
& $100.0$ & $1.01$ & $31.33$ & $0.0409$ & $1961$\\
\hdashline
\multirow{2}{4em}{${\rho=0.01}$} & $102.0$ & $1.01 $ & $32.04 $ & $0.0405 $ & $2040 $ \\
& $102.1 $ & $1.01 $ & $ 32.08$ & $0.0405 $ & $2042$ \\
\hdashline
\multirow{2}{4em}{${\rho=0.03}$} & $102.0$ & $0.99$ & $33.53 $ & $0.0402 $ & $2040 $  \\
& $101.9$ & $0.99$ & $33.42$ & $0.0402$ & $2038$\\
\hline
\end{tabularx}
\caption{Comparative Statics with respect to the scale parameter $\lambda$ and $\rho$. For each parameter, the upper row recalls the equilibrium values for fixed overall emissions, while the lower row states the values in the social welfare maximizing equilibrium. }
\label{ECP:Table:CompStatWelfareLambdaRho}
\end{table}

Next, we discuss the effect of a shift in the scale parameter $\lambda$, which measures the carbon intensity of production. Table \ref{ECP:Table:CompStatWelfareLambdaRho} summarizes our findings, where we compare the equilibrium values from the previous sensitivity analysis (for fixed emission target) with those resulting from the welfare optimizing equilibrium. Recall that an increased carbon intensity lead to a decreased overall output level and a lower carbon price, where the latter results from firms endogenously balancing their cost of production. \\
Concerning the welfare-maximizing equilibrium, we note that as carbon intensity rises, the regulator takes action by lowering the emission target. This action can be interpreted as a clear signal from the regulator to polluting firms, indicating consequences for inefficient production facilities. Consequently, the output of polluting firms drops, and as competition among firms intensifies, the regulator's action reverts the carbon price effect observed in the earlier sensitivity analysis. More precisely, we observe an increasing carbon price in response to the increasing carbon intensity. As a result, firms choose to invest in a cleaner technology at an earlier stage, such that the investment threshold $b$ declines and the turnover rate increases. 

Regarding the scale parameter $\rho$, we observe the following. Recall that an increase in $\rho$ increases the damage of emissions on the production on both polluting as well as carbon neutral firms. While this decreased expected profits and thus lead to less competition in the equilibrium model with fixed overall emissions, we observe that the regulator's actions mitigates this effect by lowering the emission target. Hence, the regulator acknowledges that production becomes less efficient when increasing the carbon induced damages, and penalizes polluting firms by reducing their available emissions. We observe that the latter leads to a decreased overall output of polluting firms as well as, as seen in Figure \ref{ECP:Fig:Regulator}, a lower social welfare. 

\section{Conclusion}
In this paper, we discuss a model of strategic green technology adoption within a stationary equilibrium framework of competing firms. We succeed in deriving the existence and uniqueness of an endogenous carbon price and a distribution of polluting firms, that achieve a given emission target and keep the economy within a steady state. To this end, our notion of equilibria includes a compliance/consistency condition, that equates net emissions with a predetermined emission target. Within this framework, each firm maximizes profits and chooses an abatement strategy, which is triggered by a level $b^*$ at which it is cost beneficial to switch to a carbon neutral technology. Our general formulation allows the study of several instances of this problem, including different technology shock process, profit functions and installation costs. Hence, we expect that our framework can accommodate other specific problem formulations that allow for the study of parameters not treated in this paper. \\
In our case study of  Section \ref{ECP:Section:ExplicitModel} we showcase some of the implications of our model for technology shocks given by arithmetic Brownian motions and a simple AK-model with damage function on production. We observe that pollution regulation via a specified quota (and thus, a carbon price on emissions) affects firms in multiple ways. First, we note that a more stringent regulation leads to a decreased industry output and thus to a larger competition on quantities. Even though firms are faced with a constant elasticity of demand function, which is not affected by the output of competing firms, the competition arises through the stipulated emission target that effectively limits output produced with dirty technology. As a result, the carbon price increases with less emissions available. Second, pollution regulation has a large effect on firms' abatement strategy. We observe that a carbon pricing system gives incentives to firms to switch once and for all to a carbon neutral technology, which makes them independent from the imposed market price on carbon. Clearly, firms are willing to invest earlier (and thus, for lower values of the underlying technology shock process) when the carbon price increases. Regarding the fixed parameters of the model, we observe that a slowdown of technology growth rate (of dirty technology) or faster technology growth (of carbon neutral technology) leads firms to switch earlier. Moreover, tax benefits for green firms (in form of a decreased corporate tax) as well as subsidies on the installation cost for abatement technology create financial incentives for polluting firms to invest, as reflected in a decreased trigger threshold $b^*$. In contrast, a policy that increases corporate taxes on polluting firms may yield suboptimal results, as it leads to a decreased market size, less competition and lower force to invest in green technology.   \\
The paper also allows the presence of a potential regulator, that aims to optimize its own objective. In our case study we shortly discuss possible consequences arising through the introduction of a welfare maximizing regulator, who optimally sets the emission target in order to weigh off the benefits and damages of polluting production. We find that the regulator may set an optimal emission target that mitigates or reverts the effect of some model's parameters on the carbon price and the firm's investment decision. For instance, a larger damage factor of pollution on production leads to less profits of market participants, which leads to less competition, a lower carbon price and a smaller force to become carbon neutral. The regulator mitigates this effect by lowering the emission target and forces firms to invest earlier. 

\section{Acknowledgements}
The authors would like to thank Michael Barnett, Sara Biagini, Marta Leocata, Frank Riedel, and Luca Taschini for fruitful discussions and helpful suggestions. Moreover, the authors gratefully acknowledge financial support from \textit{Deutsche Forschungsgemeinschaft} (DFG, German
Research Foundation) -- Project-ID 317210226 -- SFB 1283.

\singlespacing
\appendix 

\section{Technical Assumptions}

In this Section, we gather some of the assumptions needed for the analysis in the previous sections. 
Assumption \ref{ECP:Assumption:CoefficientsSDE} is concerned with the coefficients in the SDEs \eqref{ECP:SDE:brown}-\eqref{ECP:SDE:green}, while Assumption \ref{ECP:Assumption:ProfitFct} lists the employed assumptions on the profit functions of the polluting and carbon-neutral firms, as well as the sunk cost and the emissions of the polluting firms.

\begin{assumption}[Assumption on technology shock processes]\label{ECP:Assumption:CoefficientsSDE}
We denote $\mathbb{R}_+ = (0,\infty)$. We assume that the state space of the diffusions $Z^z$ and $\overline{Z}^z$ are given by $\mathcal{I} = (\underline{x}, \infty)$ and $\overline{\mathcal{I}} = (\underline{y}, \infty)$, where $\underline{x},\underline{y} \leq 0$. 
    The coefficients $\mu_i: \mathbb{R} \to \mathbb{R}$ and $\sigma_i:\mathbb{R} \to \mathbb{R}_+$, $i=1,2$ are such that $\mu_i \in C^1$, $\sigma_i \in C^2$ and
    \begin{align*}
        \vert \mu_i (x) - \mu_i(y) \vert \leq K_i \vert x-y\vert, \qquad \vert \sigma_i (x) - \sigma_i (y) \vert \leq h_i ( \vert x-y \vert)  
    \end{align*}
    for some $K_i>0$, $h_i :\mathbb{R}_+ \to \mathbb{R}_+$ strictly increasing, $h_i (0) = 0$ and
    \begin{align*}
        \int_{(0,\epsilon)} \frac{dx}{h_i^2 (x)} = \infty 
    \end{align*}
    for all $\epsilon>0$ and $x,y$ in $\mathcal{I}$, or $\overline{\mathcal{I}}$, respectively. 
\end{assumption}
\begin{remark}
    We highlight that, a priori, we do not specify whether zero is attainable or unattainable for the diffusions $Z$ and $\overline{Z}$. In the forthcoming analysis, we will assume that the firms exit the market whenever their technology shock process falls below the level zero. Clearly, if zero is unattainable for the diffusion, no absorption takes place (consider, e.g., a geometric Brownian motion). On the other hand, it is straightforward to modify the assumption to include an absorption in a point $\epsilon >0$.
\end{remark}
Under the previous assumption, for all $x \in \mathcal{I}$, respectively $\overline{\mathcal{I}}$, there exists $\epsilon>0$ such that 
\begin{align*}
    \int_{x-\epsilon}^{x+\epsilon} \frac{1 + \vert \mu_i (y) \vert }{\sigma_i^2 (y)} dy < + \infty, \qquad i = 1,2, 
\end{align*}
such that \eqref{ECP:SDE:brown}-\eqref{ECP:SDE:green} have a weak solution that is unique in the sense of probability law \citep[see][Chapter 5.5]{KaratzasShreve1991}. Since also those solutions are each pathwise unique by the Yamada-Watanabe's Theorem, it follows that \eqref{ECP:SDE:brown}-\eqref{ECP:SDE:green} have unique strong solutions \citep[see][Corollary 5.3.23]{KaratzasShreve1991} that are regular in the sense that any point of the interior of their respective state space can be reached in finite time with positive probability. We assume that the boundary points $+\infty$ are not attainable for neither of the two processes, i.e~cannot be reached in finite time with positive probability.

\begin{assumption}[Assumption on involved functions]\label{ECP:Assumption:ProfitFct}
We assume 
    \begin{itemize}
        \item[(i)] $\pi_1(\,\cdot\,;E_{max},\, \cdot \,) \in C^{1,1}(\mathbb{R_+} \times \mathbb{R}_+)$, $\pi_2(\,\cdot\,;E_{\max}) \in C^{1}(\mathbb{R_+})$, $c(\cdot) \in C^1(\mathbb{R})$, $e(\cdot) \in C(\mathbb{R})$;
        \item[(ii)] $0 \leq \pi_1(z; E_{\max}, c_p) \leq K_1(c_p), 0 \leq \pi_2 (z;E_{\max}) \leq K_2 $ for some $K_1(c_p), K_2 > 0$ and all $(z, E_{\max})\in \mathbb{R}_+^2$;
        \item[(iii)] $\frac{\partial}{\partial c_p} \pi_1(z; E_{max}, c_p) < 0 $ for all $z \geq 0$;
        \item[(iv)] $\lim_{c_p \to \infty} \pi_1(z,c_p) = 0$ for all $z\in \mathbb{R}_+$; 
        \item[(v)] $c(z), e(z) >0$ for all $z \in \mathbb{R}_+$, and $\vert c'(0) \vert < \infty$. 
        \item[(vi)] The distribution $\xi$ admits a density function $g(\cdot) \in C^1$. 
    \end{itemize}
\end{assumption}

    While Assumptions \ref{ECP:Assumption:ProfitFct} (i), (v), (vi)  deal with technical aspects, we can provide a theoretical foundation for assumptions (ii)-(iv). The second assumption posits that the profit function remains non-negative when considering positive values for technology $z$, overall emissions $E_{\max}$, and carbon price $c_p$. Furthermore, we assume that an increase in the carbon price, which firms must pay for each unit of emitted pollutants (which is indirectly linked to production output), adversely affects firm profits. Lastly, we presume that as the carbon price approaches infinity, firms' profits tend to zero. In simple terms, emitting firms cease production entirely when faced with an exorbitant carbon price, resulting in zero profits. This assumption implies the absence of fixed production costs in our model, although it is straightforward to extent the model in this regard.

\section{Single-firm optimal investment problem}
In this section, we provide a solution to the optimal investment problem \eqref{ECP:Def:ValueFunction} of the single firm. Before we explain on how to derive a candidate value function and present the proof to Theorem \ref{ECP:Theorem:VerThm}, we first state the following technical remark. 
\begin{remark} \label{ECP:Remark:Resolvent}
    Using arguments presented in \citet{Alvarez2007}, we are able to express the functions $\Phi_1$ and $\Phi_2$ in a purely analytical way. In the following, we let $\psi(z)$ and $\varphi(z)$ denote the increasing and decreasing, respectively, fundamental solutions to the ordinary differential equation $(\mathcal{L} - (r+\eta)) u(z) = 0$, where $\mathcal{L}$ denotes the infinitesimal generator associated to the diffusion $Z$. 
    We recall that the Green-kernel $G:\mathcal{I} \times \mathcal{I} \to \mathbb{R}_+$ of the linear diffusion $Z$ is given by 
    \begin{align*}
        G_{r+\eta}(z,y) = \int_0^\infty e^{-rt} p(t;z,y)dt = \begin{cases}
            W^{-1} \psi (z) \varphi (y), &z<y \\
            W^{-1} \psi (y) \varphi (z), &z\geq y,
        \end{cases}
    \end{align*}
    where $W$ denotes the constant Wronskian determinant of the fundamental solutions $\psi(z)$ and $\varphi(z)$, given by 
    \begin{align*}
        W = \frac{\psi'(z)}{S'(z)} \varphi (z) - \frac{\varphi'(z)}{S'(z)} \psi(z) >0, 
    \end{align*}
    and where $S'(z) = \exp (- \int (2 \mu_1 (z) / \sigma_1^2 (z)) dz$ denotes the scale density of the diffusion $Z$. It can then be shown \citep[see][]{Alvarez2007} that $\Phi_1(\cdot)$ as in \eqref{ECP:Def:Phi1Phi2} can be rewritten as
    \begin{align*}
        \Phi_1 (z) = \int_0^\infty G_{r+\eta}^{(0,\infty)} (z,y) \pi_1(y) m'(y) dy,      \end{align*}
    where $m'(y) = 2/(\sigma_1^2(y) S'(y))$ denotes the speed density of $Z$ and $G_{r+\eta}^{(0,\infty)} (z,y)$ the Green-kernel of the constrained process $Z$ that is killed at $0$, given by 
    \begin{align*}
        G_{r+\eta}^{(0,\infty)} (z,y)  = \begin{cases}
            W^{-1} \varphi (y) \psi (z,0), &z<y, \\
            W^{-1} \varphi (z) \psi(y,0),&z\geq y,
        \end{cases}
    \end{align*}
    where we let $\psi (z,0) = \psi(z) - (\psi(0)/\varphi(0)) \varphi (z)$. We note that 
    \begin{align*}
        %\lim_{a \downarrow 0} 
        \psi (z,0) = \begin{cases}
            \psi (z), &\text{if $0$ is natural or exit,} \\
            \psi(z) - \frac{\psi(0)}{\varphi(0)} \varphi (z), & \text{if $0$ is regular or entrance.}
        \end{cases}
    \end{align*}
    For a classification of boundary points we refer to \citet{BorodinSalminen2015}, Chapter $2$. Analogously, we let $\overline{\psi}(z)$ and $\overline{\varphi}(z)$ denote the fundamental solutions to the ordinary differential equation $(\overline{\mathcal{L}} -(r+ \eta)) u(z) = 0$, where $\overline{\mathcal{L}}$ is the infinitesimal generator associated to the diffusion $\overline{Z}$ of \eqref{ECP:SDE:green}. We can proceed as above and obtain
    \begin{align*}
         \Phi_2 (z) = \int_0^\infty \overline{G}_{r+\eta}^{(0,\infty)} (z,y) \pi_2(y) \overline{m}'(y) dy, 
    \end{align*}
    where $\overline{W}, \overline{S}'(z), \overline{m}'(z), \overline{G}_{r+\eta}^{(0,\infty)} (x,y)$ denote the Wronskian, scale density, speed density and Green-kernel of the constrained process $\overline{Z}$, respectively.
    %\begin{align*}
    %    \Phi_i (z) = W^{-1} \Big( \varphi (z) \int_0^z \psi(y,0) \pi_i (y) m '(y)dy + \psi(z,0) \int_z^\infty \varphi(y) \pi_i(y) m'(y)dy \Big)
    %\end{align*}
\end{remark}

In the following, we derive a suitable candidate for the value function $v$ of \eqref{ECP:Def:ValueFunction}, which we then verify to be the true value function. To begin with, and via standard techniques, we associate the value function with a variational inequality of the form
\begin{align}\label{ECP:Equ:VariationalInequ}
    \max \big\{ (\mathcal{L} -(r+ \eta) w(z) + \pi_1 (z) , ~ \Phi_2 (z) - c(z) - w(z) \big\} = 0,
\end{align}
with boundary condition $w(0)=0$ and where $\mathcal{L} := \frac{1}{2}\sigma_1^2 (z) \partial_{zz} + \mu_1(z) \partial_z$ denotes the infinitesimal generator associated to the diffusion $Z$ of \eqref{ECP:SDE:brown}. Intuitively, equation \eqref{ECP:Equ:VariationalInequ} arises through the dynamic programming approach and 
reflects the two available strategies of the firm at any given point in time: If the firm chooses to invest in the carbon neutral technology, it receives a payoff of $\Phi_2 - c$. Otherwise, the firm continues producing as a polluting firm, receives the running profit $\pi_1$, and holds on to the option to invest. Equation \eqref{ECP:Equ:VariationalInequ} formalizes this heuristic argument and displays that the firm chooses the maximum between theses two rewards. As usual in optimal stopping problems, the state space then splits in two distinct regions: The \textit{waiting} and \textit{stopping region}. While it is optimal to invest immediately in the latter region, the firm should delay investment in the waiting region. In the following, we follow a guess-and-verify approach and conjecture that firms will invest (and thus become carbon neutral) whenever their technology level is sufficiently large. Hence, we guess that there exists a threshold $b>0$ such that the stopping time
\begin{align}\label{ECP:Def:StopTime}
    \tau_b := \inf \{ t \geq 0: ~ Z_t \geq b \}
\end{align}
is optimal. 
Accordingly, we can relate the above variational inequality \eqref{ECP:Equ:VariationalInequ} to the following \textit{free-boundary problem}
\begin{align}\label{ECP:Equ:FreeBoundaryProblem}
    \begin{cases}
        (\mathcal{L} -(r+\eta)w(z) + \pi_1 (z) = 0, & 0<z<b, \\
        (\mathcal{L} -(r+\eta)w(z) + \pi_1 (z) \leq 0, & b \leq z, \\
        w(z) = \Phi_2 (z) - c(z), & b \leq z,\\
        w(z) \geq \Phi_2 (z) - c(z), & 0< z <b, \\
        w(0) = 0, & 
    \end{cases}
\end{align}
where we separated the values of technology into the waiting region $\mathbb{W} = (0,b)$ and the stopping region $\mathbb{S} = [b,\infty)$.
We solve the free-boundary problem \eqref{ECP:Equ:FreeBoundaryProblem} by first recalling that any solution to the ODE $(\mathcal{L} - (r+ \eta))w(z) + \pi_1(z) = 0$ is given by 
\begin{align*}
    w(z) = A \psi (z) + B \varphi (z) + \Phi_1 (z), 
\end{align*}
where $\psi(\cdot)$ and $\varphi(\cdot)$ are as in Remark \ref{ECP:Remark:Resolvent}.
The boundary condition implies $A \psi (0) + B \varphi ( 0) = 0$, such that 
\begin{align*}
    w(z) =  A \frac{\psi (z) \varphi (0) - \varphi (z)\psi (0)}{\varphi (0)} + \Phi_1 (z) =: A \psi (z,0) + \Phi_1(z),
\end{align*}
where $\psi(z,0)$ is defined as in Remark \ref{ECP:Remark:Resolvent}. 
For the ease of notation, we now let 
\begin{align*}
    G(z) := \Phi_2(z)- \Phi_1(z) - c(z),
\end{align*}
and, by imposing smooth-fit and smooth-pasting conditions at the free boundary $b$, we obtain that the \textit{candidate value function} takes the form
\begin{align}\label{ECP:Def:CandValueFunction}
    w(z) := \begin{cases}
        0, & z \leq 0, \\
       \Phi_1(z) + \frac{G(b)}{\psi(b,0)} \psi(z,0), & 0<z<b, \\
        \Phi_2(z) - c(z), & z \geq b,
    \end{cases}
\end{align}
and the optimal stopping threshold $b$ is given by the solution -- provided that it exists -- to the equation
\begin{align*}
    %G(b) \frac{\psi(0)\varphi'(b) - \varphi(0) \psi'(b)}{\psi(0) \varphi(b) - \varphi (0) \psi(b)} = G'(b) \\
    %\Leftrightarrow  
    G'(b) [\psi(0) \varphi(b) - \varphi (0) \psi(b)] - G(b)[ \psi(0)\varphi'(b) - \varphi(0) \psi'(b)] = 0.
\end{align*}
We define 
\begin{align*}
    A (z) := \frac{G'(z) [\psi(0) \varphi(z) - \varphi (0) \psi(z)] - G(z)[ \psi(0)\varphi'(z) - \varphi(0) \psi'(z)]}{S'(z)},
\end{align*}
where $S'(z)$ denotes the scale density of the diffusion $Z$. Furthermore, as in Remark \ref{ECP:Remark:Resolvent}, we let $m'(z) = 2/(\sigma^2 (z) S'(z))$ denote the speed measure density. We now search for a point $b$ such that $A(b) = 0$. Since by direct computations one has 
\begin{align}\label{ECP:Equ:DerA(x)}
    \frac{d}{d z} A(z) 
    %&= m'(x) \Big( (\psi(0)\varphi(x) - \varphi (0) \psi (x)) \big( \mathcal{L} - (r  + \eta) \big) G(x) - G(x) (\mathcal{L} - (r- \eta)) (\psi (0) \varphi (x) - \varphi(0) \psi (x)) \Big) \\
    &= m'(z) \big( \psi (0) \varphi (z) - \varphi (0) \psi (z) \big) \big( \mathcal{L} - (r + \eta) \big) G(z), 
\end{align}
we can write 
\begin{align*}
    A(b) = A(0) + \int_0^b \frac{d}{d z} A(z) dz,
\end{align*}
and notice that, under Assumption \ref{ECP:Assumption:ProfitFct}, we have
\begin{align*}
    A(0) &= G'(0)\frac{\varphi(0) \psi (0) - \varphi (0) \psi (0)}{S'(0)} +  G(0) \frac{\varphi(0) \psi '(0) - \varphi '(0) \psi (0)}{S'(0)} \\
    &= \big( \Phi_2(0) - \Phi_1(0) - c(0) \big) W = -c(0) W < 0, 
\end{align*}
where the latter inequality follows from Assumption \ref{ECP:Assumption:ProfitFct}.
We are now ready to prove Theorem \ref{ECP:Theorem:VerThm}. 

\subsection{Proof of Theorem \ref{ECP:Theorem:VerThm}}\label{ECP:Appendix:VerificationTheorem}

Proof of (i). We establish the result in three steps. \par
\vspace{0.15cm}
\textit{Step 1.} First, we prove that there exists a unique point $b \in \mathbb{R}_+$ such that $A(b) = 0.$ To this end, we recall that 
    \begin{align*}
         A(b) = A(0) + \int_0^b \frac{d}{d z} A(z) dz = - c(0) W  + \int_0^b \frac{d}{d z} A(z) dz.
    \end{align*}
    Since $-c(0)W<0$ and $A'(z) > 0$ for all $z> \tilde{z}$ under Assumption \ref{ECP:Assumption:(L-r)G}, it is sufficient to prove that $\lim_{b \to \infty} A(b) = + \infty.$ Straightforward calculations, upon employing the mean value theorem for some point $\xi \in (\tilde{z}, b)$, yield
    \begin{align*}
        &\lim_{b \to \infty}  A(b) = - c(0) W  + \lim_{b \to \infty} \int_0^b \frac{d}{d z} A(z) dz \\
        &= - c(0) W  + \int_0^{\tilde{z}} \frac{d}{d z} A(z) dz + \lim_{b \to \infty} \int_{\tilde{z}}^b m'(z) \big( \psi (0) \varphi (z) - \varphi (0) \psi (z) \big) \big( \mathcal{L} - (r + \eta) \big) G(z) dz \\
        &= - c(0) W  + \int_0^{\tilde{z}}  \frac{d}{d z} A(z) dz \\
        &\hspace{3cm} + \lim_{b \to \infty} \frac{\big( \mathcal{L}-(r+\eta) \big)G(\xi)}{r + \eta} \int_{\tilde{z}}^b (r + \eta) m'(z) ( \psi (0) \varphi (z) - \varphi (0) \psi (z)) dz \\
        &= - c(0) W  + \int_0^{\tilde{z}}  \frac{d}{d z} A(z) dz \\
        &\hspace{3cm}+ 
        \lim_{b \to \infty} \frac{(\mathcal{L}-(r+\eta))G(\xi)}{r + \eta}
        \bigg( \psi(0) \Big(\frac{\varphi'(b)}{S'(b)} - \frac{\varphi'(\tilde{z})}{S'(\tilde{z})}\Big)- \varphi(0) \Big(\frac{\psi'(b)}{S'(b)} - \frac{\psi'(\tilde{z})}{S'(\tilde{z})}\Big) \bigg) \\
        &\to + \infty,
    \end{align*}
    and the latter follows from Assumption \ref{ECP:Assumption:ProfitFct}, $(\mathcal{L}-(r+\eta))G(\xi) <0$ (since $\tilde{z}< \xi)$, and $\varphi'(b)/S'(b) \downarrow 0$ as well as $\psi'(b)/S'(b) \uparrow +\infty$  \citep[see for example][Chapter $2$]{BorodinSalminen2015}. \par 
    \vspace{0.15cm}
\textit{Step 2.} Next, we show that the candidate
value function $w(\cdot)$ of \eqref{ECP:Def:CandValueFunction} solves the variational inequality      \eqref{ECP:Equ:VariationalInequ}. 
    Let $b\in \mathbb{R}_+$ denote the unique solution to $A(\cdot) = 0$, as determined in Step 1. By construction, $(\mathcal{L} - (r+ \eta))w(z) + \pi_1(z) = 0$ on $(0,b)$, while $w(z) = \Phi_2(z) -c(z)$ on $(b,\infty)$. In order to show that $w$ solves the variational inequality on $\mathbb{R_+}$, it thus remains to show (i) $(\mathcal{L} - (r+ \eta))w(z) + \pi_1(z) \leq 0$ on $(b,\infty)$ and (ii) $w(z) \geq \Phi_2(z) -c(z)$ on $(0,b)$. \\
    Regarding (i), we recall that $w(z) = \Phi_2(z) - c(z) = G(z) + \Phi_1(z)$ on $(b, \infty)$. It follows that 
    \begin{align*}
        (\mathcal{L} - (r + \eta))w(z) + \pi_1(z) = (\mathcal{L} - (r + \eta))G(z) < 0, 
    \end{align*}
    where the latter inequality follows from Assumption \ref{ECP:Assumption:(L-r)G} and by construction, since $b>\tilde{z}$. \\
    Regarding (ii), we recall that $w(z) = \Phi_1(z) + (G(b)/\psi(b,0)) \psi(z,0)$ on $(0,b)$. Simple calculations reveal that $w(z) \geq \Phi_2(z) -c(z)$ is equivalent to 
    \begin{align*}
        \frac{G(b)}{\psi (0) \varphi (b) - \varphi (0) \psi(b)} \leq \frac{G(z)}{\psi (0) \varphi (z) - \varphi (0) \psi(z)},
    \end{align*}
and consequently, it is sufficient to prove that $b$ is a local minimum of the function $z \mapsto G(z)/\allowbreak (\psi (0) \varphi (z)  - \varphi (0) \psi(z))$ on $(0,b).$ We compute 
\begin{align*}
    \bigg( \frac{G(z)}{\psi (0) \varphi (z) - \varphi (0) \psi(z)} \bigg)' = A(z) \frac{S'(z)}{(\psi (0) \varphi (z) - \varphi (0) \psi(z))^2},
\end{align*}
which is zero for $z=b$ by construction. Furthermore,
\begin{align*}
  \bigg( \frac{G(z)}{\psi (0) \varphi (z) - \varphi (0) \psi(z)} \bigg)'' &\Big\vert_{z=b} \\
  &= \bigg( A'(b) \frac{S'(b)}{(\psi (0) \varphi (b) - \varphi (0) \psi(b))^2} + A(b) \Big( \frac{S'(b)}{\psi (0) \varphi (b) - \varphi (0) \psi(b)} \Big)' \bigg) \\
  &= A'(b) \frac{S'(b)}{(\psi (0) \varphi (b) - \varphi (0) \psi(b))^2}  > 0,
\end{align*}
where the latter inequality follows from Assumption \ref{ECP:Assumption:(L-r)G} and $b>\tilde{z}$. Hence, $b$ is a local minimum of the function $z \mapsto G(z)/\allowbreak (\psi (0) \varphi (z)  - \varphi (0) \psi(z))$ on $(0,b)$ and the claim follows. \par
\vspace{0.15cm}
\textit{Step 3.} Last, we verify that the candidate value function $w$ indeed coincides with the true value function $v$ of \eqref{ECP:Def:ValueFunction} and that the threshold $b$ triggers the optimal stopping time $\tau_b$ as in \eqref{ECP:Def:StopTime}. 
    Let $n \in \mathbb{N}$ and define $\sigma_n := \inf \{ t\geq 0: Z_t^z \geq n \}.$ We let  $\tau_n = \tau \wedge \sigma_n$ for any stopping time $\tau$ of the Brownian filtration. Due to our construction of the function $w$, we can employ It\^o's formula to obtain
    \begin{align*}
        e^{-(r+\eta) (\tau_n \wedge \gamma_1)} w(Z_{\tau_n \wedge \gamma_1}^z) = w(z) + \int_0^{\tau_n \wedge \gamma_1} e^{-(r+\eta) s} (\mathcal{L} - (r+\eta)) w(Z_s^z) \one_{\{ Z_s^z \neq b\}} ds + M_{\tau_n \wedge \gamma_1} 
    \end{align*}
    where $\gamma_1$ is defined as in \eqref{ECP:Def:Gamma12} and $M_t$ denotes the stochastic integral 
    \begin{align*}
        M_t = \int_0^t e^{-(r+\eta)s} \sigma_1 (Z_s^z) w' (Z_s^z) dW_s, \qquad t\geq 0.
    \end{align*}
    Due to the regularity of $w(\cdot)$ and $\sigma_1(\cdot)$ we observe $\mathbb{E} [M_{\tau_n \wedge \gamma_1}] = 0$, such that taking expectations yields
    \begin{align*}
        \mathbb{E} \big[  e^{-(r+\eta) (\tau_n \wedge \gamma_1)} w(Z_{\tau_n \wedge \gamma_1}^z) \big] = w(z) + 
        \mathbb{E} \Big[ \int_0^{\tau_n \wedge \gamma_1} e^{-(r+\eta) s} (\mathcal{L} - (r+\eta)) w(Z_s^z) ds 
        \Big],
    \end{align*}
    where we used $\mathbb{P} [ Z_s^z =b]=0$. Since $w$ solves the variational inequality \eqref{ECP:Equ:VariationalInequ}, as proven in Step 2., we notice that $w(z) \geq \Phi_2(z) - c(z)$ as well as $(\mathcal{L} - (r+ \eta))w(z) \leq - \pi_1(z)$ a.e. It follows that 
    \begin{align*}
        w(z) \geq \mathbb{E} \Big[ 
        \int_0^{\tau_n \wedge \gamma_1} e^{-(r+\eta) s} \pi_1 (Z_s^z) ds + e^{-(r+\eta) \tau_n} \big( \Phi_2 (Z_{\tau_n}^z) - c(Z_{\tau_n}^z) \big) \one_{ \{ \tau_n < \gamma_1 \} }
        \Big],
    \end{align*}
    where we employed the equality $w(Z^z_{\gamma_1}) = w(0) =0$, which follows by construction. Letting $n \to \infty$, invoking the dominated convergence theorem due to Assumption \ref{ECP:Assumption:ProfitFct}, we obtain
    \begin{align*}
        w(z) \geq \mathbb{E} \Big[ 
        \int_0^{\tau \wedge \gamma_1} e^{-(r+\eta) s} \pi_1 (Z_s^z) ds + e^{-(r+\eta) \tau} \big( \Phi_2 (Z_{\tau}^z) - c(Z_{\tau}^z) \big) \one_{ \{ \tau < \gamma_1 \} }
        \Big] = J(z, \tau),
    \end{align*}
    for all $\tau \in \mathcal{T}$, and thus $w(z) \geq v(z)$ for all $z \in \mathbb{R}_+$. Repeating the previous steps, but choosing the stopping time $\tau_b$ of \eqref{ECP:Def:StopTime} with $b$ determined via the equation $A(\cdot) = 0$, we obtain 
    \begin{align*}
        w(z) &= J(z, \tau_b) 
        \leq \sup_{\tau} J(z, \tau) = v(z),
    \end{align*}
    for all $z \in \mathbb{R}_+$. It follows that $w = v$ and $\tau_b$ is an optimal stopping time for problem \eqref{ECP:Def:ValueFunction}.  \par 
    \vspace{0.25cm}
    Proof of (ii). We prove the claims separately. Without the risk of confusion, and where necessary, we denote $A(b(c_p),c_p)=A(b),~\Phi_1(z,c_p) = \Phi_1(z)$ as well as $G(z, c_p) = G(z)$, in order to highlight their dependency on $c_p>0$. \par 
    
    First, we show that the function $c_p \mapsto b(c_p)$ is strictly decreasing, as well as that the limit $\lim_{c_p\to\infty} b(c_p)=: b_\infty >0 $ exists. Since        \begin{align}\label{ECP:Equ:Derb(cp)}
            b'(c_p) = \frac{\frac{\partial}{\partial c_p} A(b(c_p),c_p)}{\frac{\partial}{\partial b} A(b(c_p),c_p)},
        \end{align}
        and $\frac{\partial}{\partial b} A(b(c_p),c_p) > 0$ due to \eqref{ECP:Equ:DerA(x)} and Assumption \ref{ECP:Assumption:(L-r)G}, it is thus left to study the sign of $\frac{\partial}{\partial c_p} A(b(c_p),c_p)$. Simple calculations yield, for any $b>0$,
        \begin{align*}
            \frac{\partial}{\partial c_p} A(b,c_p) = \frac{ G_{b c_p} (b,c_p) [ \varphi (b) \psi(0) - \varphi (0) \psi(b)] - G_{c_p} (b,c_p) [\varphi'(b)\psi(0) - \varphi(0)\psi'(b)]}{S'(b)}.
        \end{align*}
        Moreover, since $G(b,c_p) = \Phi_2(b) - \Phi_1(b,c_p) -c(b)$, where the first and third term are clearly independent of $c_p$, we have
        \begin{align*}
            G_{c_p} &(b,c_p)\\
            &= 
            %- \frac{\partial}{\partial c_p} \Phi_1(b,c_p) \\ &=
           - W^{-1} \Big( \varphi (b) \int_0^b \psi(z,0) \frac{\partial}{\partial c_p} \pi_1(z,c_p)m'(z)dz + \psi(b,0)\int_b^\infty \varphi(z) \frac{\partial}{\partial c_p} \pi_1(z,c_p)m'(z)dz \Big), \\
           G_{b c_p} &(b,c_p) \\ 
            &= - W^{-1} \Big( \varphi' (b) \int_0^b \psi(z,0) \frac{\partial}{\partial c_p} \pi_1(z,c_p)m'(z)dz 
            + \varphi (b) \psi(b,0) \frac{\partial}{\partial c_p} \pi_1 (b, c_p) m'(b) \\
            &\hspace{1cm} + \psi '(b,0)\int_b^\infty \varphi(z) \frac{\partial}{\partial c_p} \pi_1(z,c_p)m'(z)dz - \varphi (b) \psi(b,0) \frac{\partial}{\partial c_p} \pi_1 (b, c_p) m'(b) \Big) \\
            &= - W^{-1} \Big( \varphi' (b) \int_0^b \psi(z,0) \frac{\partial}{\partial c_p} \pi_1(z,c_p)m'(z)dz + \psi '(b,0)\int_b^\infty \varphi(z) \frac{\partial}{\partial c_p} \pi_1(z,c_p)m'(z)dz \Big).
        \end{align*}
        The numerator of \eqref{ECP:Equ:Derb(cp)} then rewrites as 
        \begin{align*}
              G_{b c_p} (b,c_p) [\varphi(b)\psi(0) &- \varphi(0)\psi(b)] -  G_{c_p} (b,c_p) [\varphi'(b)\psi(0)-\varphi(0)\psi'(b)] \\
             &= W^{-1} \varphi(0)[\varphi'(b)\psi(b)-\varphi(b)\psi'(b)] \int_0^b \psi(z,0)\frac{\partial}{\partial c_p} \pi_1(z,c_p) m'(z) dz  \\
             &= - S'(b) \varphi(0) \int_0^b \psi(z,0)\frac{\partial}{\partial c_p} \pi_1(z,c_p) m'(z) dz,
        \end{align*}
        and it follows that 
        \begin{align*}
            \frac{\partial}{\partial c_p} A(b,c_p) = - \varphi (0) \int_0^b \psi(z,0) \frac{\partial}{\partial c_p} \pi_1(z,c_p) m'(z)dz >0,
        \end{align*}
        where the latter inequality follows from $m'(\cdot)>0$ and Assumption \ref{ECP:Assumption:ProfitFct}. 

        Regarding the limiting behaviour of $b(c_p)$ as $c_p \to \infty$ , we recall Assumption \ref{ECP:Assumption:ProfitFct} (iv), i.e. $\lim_{c_p \to \infty} \pi_1 (z,c_p) = 0$. It follows that $b_\infty := \lim_{c_p \to \infty} b(c_p)$ is the solution to 
        \begin{align}\label{ECP:Equ:A(binfty)}
            A(b_\infty) = - c(0) W + \int_0^{b_\infty} m'(z) \big( \psi(0) \varphi(z)  - \psi (z) \varphi(0) \big) \big( \mathcal{L} - (r+ \eta)\big) \big(\Phi_2 (z) - c(z)\big)dz. 
        \end{align}
        Notice that Assumption \ref{ECP:Assumption:(L-r)G} is still satisfied, such that $b_\infty$ is uniquely determined via \eqref{ECP:Equ:A(binfty)}. Moreover, it follows that $b_\infty > 0$. \par 

        Next, we prove that the map $c_p \mapsto v(z;c_p)$ is strictly decreasing on $z \in (0,b(c_p))$ and constant on $(b(c_p),\infty)$. 

    Clearly, $v_{c_p} (z;c_p) = 0$ for $z\geq b(c_p)$. For $z\in (0,b(c_p))$, we compute
    \begin{align*}
        v_{c_p}(z,c_p) &=       \frac{\partial}{\partial c_p} \Big( \Phi_1 (z,c_p) + G(b(c_p),c_p) \frac{\psi (z) \varphi (0) - \varphi (z) \psi (0)}{\psi (b(c_p)) \varphi (0) - \varphi (b(c_p)) \psi (0)} \Big) \\
        &= \frac{\partial }{\partial c_p} \Phi_1 (z,c_p) + \Big( G_{c_p} (b(c_p),c_p) + b'(c_p) G_b (b(c_p),c_p) \Big) \frac{\psi(z)\varphi(0) - \varphi(z)-\psi(0)}{\psi(b(c_p))\varphi(0) - \varphi(b(c_p))\psi(0)} \\
        &\hspace{0.3cm} + G(b(c_p),c_p) b'(c_p)\big( \psi (0) \varphi'(b(c_p)) - \varphi(0) \psi'(b(c_p)) \big) \frac{\psi(z)\varphi(0) - \varphi(z)-\psi(0)}{(\psi(b(c_p))\varphi(0) - \varphi(b(c_p))\psi(0))^2}.
    \end{align*}
    Recall that
    \begin{align*}
        G(b(c_p),c_p) \frac{\psi (0) \varphi'(b(c_p)) - \varphi(0) \psi'(b(c_p))}{\varphi (b(c_p)) \psi(0) - \psi(b(c_p))\varphi(0)}  = G_b (b(c_p),c_p),
    \end{align*}
    and notice that $G_{c_p} (z,c_p) = - \frac{\partial}{\partial c_p} \Phi_1 (z,c_p)$.
    It follows that
    \begin{align*}
         v_{c_p}(z,c_p) &= \frac{\partial}{\partial c_p}  \Phi_1 (z,c_p) - \frac{\partial}{\partial c_p} \Phi_1 (b(c_p),c_p) \frac{\psi(z)\varphi(0) - \varphi(z)\psi(0)}{\psi(b(c_p))\varphi(0) - \varphi(b(c_p))\psi(0)},
    \end{align*}
    and the latter is negative if and only if 
    \begin{align*}
        \frac{\frac{\partial}{\partial c_p} \Phi_1(b(c_p),c_p) }{\psi(b(c_p))\varphi(0) - \varphi(b(c_p))\psi(0)} > 
        \frac{\frac{\partial}{\partial c_p} \Phi_1(z,c_p) }{\psi(z)\varphi(0) - \varphi(z)\psi(0)}.
    \end{align*}
    It is thus sufficient to prove that
    \begin{align*}
        &\frac{\partial}{\partial z}\Big(\frac{\frac{\partial}{\partial c_p} \Phi_1(z,c_p) }{\psi(z)\varphi(0) - \varphi(z)\psi(0)} \Big) \\
        &\qquad \quad = \frac{\frac{\partial^2}{\partial z \partial c_p} \Phi_1(z,c_p) [\psi(z)\varphi(0) - \varphi(z)\psi(0)] - \frac{\partial}{\partial c_p} \Phi_1(z,c_p) [\psi'(z)\varphi(0)-\varphi'(z)\psi(0)]}{(\psi(z)\varphi(0) - \varphi(z)\psi(0))^2} >0,
    \end{align*}
    and the latter inequality follows from employing similar arguments as in the proof of the monotonicity of $c_p \mapsto b(c_p)$. \qed 
    
\section{Proof of Proposition \ref{ECP:Proposition:StatEquilibrium}}\label{ECP:Appendix:ProofStatEqui}
\noindent    We derive the desired result in a number of steps, as described in the paragraph stated before Proposition \ref{ECP:Proposition:StatEquilibrium}. \par
\vspace{0.15cm}
    \textit{Step 1.} We begin by determining the unique equilibrium carbon price via the entry condition \eqref{ECP:Equ:EntryCondition}. We recall that $c_p \to v(z; c_p, E_{max})$ is decreasing (see Theorem \ref{ECP:Theorem:VerThm}), and hence, so is $c_p \to \int_{\underline{z}}^{\overline{z}} v(z; c_p, E_{max}) \xi (dz)$. We distinguish two cases.
    
    (i) $b_\infty > \underline{z}$. Recall that Assumptions \ref{ECP:Assumption:EntryCost} implies 
    \begin{align*}
       \int_{\underline{z}}^{\overline{z}} v(z; 0, E_{max}) \xi (dz) > c_e > \int_{\underline{z}}^{\overline{z}} \lim_{c_p \to \infty} v(z; c_p, E_{max}) \xi (dz). 
    \end{align*}
    Clearly, by the intermediate value theorem, there exists a unique $c_p^* \in (0,\infty)$ such that the entry condition is satisfied. Moreover, $b(c_p^*) \geq b_\infty > \underline{z}$ due to Theorem \ref{ECP:Theorem:VerThm}, which implies positive entry in the market.

    (ii) $b_\infty \leq \underline{z}$. Assumption \ref{ECP:Assumption:EntryCost} guarantees 
     \begin{align*}
       \int_{\underline{z}}^{\overline{z}} v(z; 0, E_{max}) \xi (dz) > c_e > \int_{\underline{z}}^{\overline{z}} v(z; \overline{c}_p, E_{max}) \xi (dz),
    \end{align*}
    where $\overline{c}_p$ is chosen such that $b(\overline{c}_p) = \underline{z}$. Again, via the intermediate value theorem, there exists a unique $c_p^* \in (0,\overline{c}_p)$ such that the entry condition is satisfied. Moreover, $b(c_p^*) \geq b(\overline{c}_p) = \underline{z}$, and hence, there is positive entry. \par 
    \vspace{0.15cm}
    \textit{Step 2.} Next, we solve for the stationary distribution of polluting firms. As described, we start by deriving its scaled density $f^*$ up to the scale factor $N^*$, i.e.~the entry rate, which is derived below in the third step of this proof. In the following, we set $b^* = b(c_p^*)$. 
    For the ease of notation, we let 
    \begin{align*}
        \tilde{\mathcal{L}} = b(z) \frac{\partial^2}{\partial z^2} + a(z) \frac{\partial }{\partial z}
    \end{align*}
    with $b(z) := \sigma^2_1 (z) /2,~ a(z) := (\sigma^2_1 (z) )_z  - \mu_1 (z) $ and $r(z) := \eta + (\mu_1 (z))_z  - (\sigma^2_1 (z))_{zz} / 2$. It follows that we can write \eqref{ECP:Equ:ODENoEntry} and \eqref{ECP:Equ:ODEEntry} equivalently as $(\tilde{\mathcal{L}} - r(z))f(z) = 0$ and $(\tilde{\mathcal{L}} - r(z))f(z) + g(z) = 0$, respectively. It is then straightforward to see that the scaled density $f$ should satisfy one of the following systems of equations
\begin{align*}
(\textrm{I})&
    \begin{cases}
        (\tilde{\mathcal{L}} - r(z))f(z) = 0,  &0 < z < \underline{z}, \\ 
        (\tilde{\mathcal{L}} - r(z))f(z) + g(z) = 0, &\underline{z} < z < b^*, 
    \end{cases}
   \\
    (\textrm{II})&
     \begin{cases}
        (\tilde{\mathcal{L}} - r(z))f(z) = 0,  &0 < z < \underline{z}, \\ 
        (\tilde{\mathcal{L}} - r(z))f(z) + g(z) = 0, &\underline{z} < z < \overline{z}, \\
        (\tilde{\mathcal{L}} - r(z))f(z) = 0,  &\overline{z} < z < b^*,
    \end{cases}
\end{align*}
where we distinguished the cases (\textrm{I}) $b^* \leq \overline{z}$ and (\textrm{II}) $b^* > \overline{z}$, respectively. 
We notice that the equation $(\tilde{\mathcal{L}} - r(z)) f(z) = 0$ admits two fundamental strictly positive solutions $\tilde{\psi} (\cdot)$ and $\tilde{\varphi} (\cdot)$, with $\tilde{\psi} (\cdot)$ being strictly increasing and $\tilde{\varphi} (\cdot)$ being strictly decreasing \citep[see][Chapter 2, Section 10]{BorodinSalminen2015}. Moreover, our assumption on the entry distribution guarantees that the equation $(\tilde{\mathcal{L}} - r(z)) f(z) + g(z) = 0$ admits a particular solution, which we denote by $G(\cdot)$. 
Using these, we can write the general solution to the systems (\textrm{I}) and (\textrm{II}), and hence our scaled density $f$, as 
\begin{align}
    (\textrm{I}) \qquad 
    f(z) &:=
    \begin{cases}
        A_1 \tilde{\psi} (z) + A_2 \tilde{\varphi} (z), &0<z<\underline{z} \\
        B_1 \tilde{\psi} (z) + B_2 \tilde{\varphi} (z) + G(z), &\underline{z} < z < b^* 
    \end{cases}
    \label{ECP:Equ:Densf(z)I}\\
    (\textrm{II}) \qquad 
     f(z) &:= \begin{cases}
        C_1 \tilde{\psi} (z) + C_2 \tilde{\varphi} (z), &0<z<\underline{z} \\
        D_1 \tilde{\psi} (z) + D_2 \tilde{\varphi} (z) + G(z), &\underline{z} < z < \overline{z} \\ 
        E_1 \tilde{\psi} (z) + E_2 \tilde{\varphi} (z), & \overline{z} < z < b^*, 
    \end{cases}
     \label{ECP:Equ:Densf(z)II}
\end{align}
for some constants $A_i, B_i, C_i, D_i, E_i, i=1,2$ to be determined.
We can solve for the constants by imposing the usual boundary conditions to ensure smoothness of the function $f$, given by $f(0) = 0,\, f(b^*) = 0, \, f(\underline{z} - ) = f(\underline{z}+), \, f'(\underline{z} - ) = f'(\underline{z}+)$ and, in the case $b^* > \overline{z}$, the additional conditions $f(\overline{z} - ) = f(\overline{z}+), \, f'(\overline{z} - ) = f'(\overline{z}+)$. Here, the first two conditions reflect that firms either exit due to their technology shock process either falling below zero, or exceeding the threshold $b^*$. Hence, there do not exist any polluting firms with a technology level outside of the set $(0,b^*)$. 
In both cases, we are able to derive unique solutions to the system of equations. In case (\textrm{I}), the constants are given by 
\begingroup
\allowdisplaybreaks
\begin{align*}
    A_1^* &= \frac{\tilde{\varphi}(0) \big(
    G(\underline{z}) [ \tilde{\psi}(b^*) \tilde{\varphi}'(\underline{z}) - \tilde{\psi}'(\underline{z}) \tilde{\varphi} (b^*)] 
    + G(b^*) [\tilde{\psi} '(\underline{z}) \tilde{\varphi} (\underline{z}) - \tilde{\psi}(\underline{z}) \tilde{\varphi}'(\underline{z})] \big)
    }{
    [\tilde{\psi}'(\underline{z}) \tilde{\varphi}(\underline{z}) - 
    \tilde{\psi}(\underline{z}) \tilde{\varphi}'(\underline{z})] \times 
    [\tilde{\psi}(0) \tilde{\varphi} (b^*) - \tilde{\psi} (b^*) \tilde{\varphi} (0)]
    } \\
    &\hspace{1.5cm} +
    \frac{
    \tilde{\varphi}(0) G'(\underline{z})  [
    \tilde{\psi}(\underline{z}) \tilde{\varphi} (b^*) - \tilde{\psi}(b^*) \tilde{\varphi} (\underline{z})
    ]
    }{
    [\tilde{\psi}'(\underline{z}) \tilde{\varphi}(\underline{z}) - 
    \tilde{\psi}(\underline{z}) \tilde{\varphi}'(\underline{z})] \times 
    [\tilde{\psi}(0) \tilde{\varphi} (b^*) - \tilde{\psi} (b^*) \tilde{\varphi} (0)]
    } \\
    A_2^* &= - A_1^* \frac{\tilde{\psi}(0)}{\tilde{\varphi}(0)} \\
    B_1^* &= A_1^* \frac{
    \tilde{\varphi}(b) [ \tilde{\psi}(\underline{z}) \tilde{\varphi}(0) - 
    \tilde{\psi}(0) \tilde{\varphi}(\underline{z})]
    }{
    \tilde{\varphi} (0)[ \tilde{\psi}(\underline{z}) \tilde{\varphi} (b^*) - 
    \tilde{\psi}(b^*) \tilde{\varphi}(\underline{z})
    ]
    } + 
    \frac{
    G(b^*) \tilde{\varphi}(\underline{z}) - G(\underline{z}) \tilde{\varphi}(b^*)
    }{
    \tilde{\psi}(\underline{z}) \tilde{\varphi}(b^*) - \tilde{\psi}(b^*) \tilde{\varphi}(\underline{z}) 
    } \\
    B_2^* &= A_1^* \frac{
    \tilde{\psi}(b) [ \tilde{\psi}(\underline{z}) \tilde{\varphi}(0) - 
    \tilde{\psi}(0) \tilde{\varphi}(\underline{z})]
    }{
    \tilde{\varphi} (0)[ \tilde{\psi}(\underline{z}) \tilde{\varphi} (b^*) - 
    \tilde{\psi}(b^*) \tilde{\varphi}(\underline{z})
    ] 
    } - 
    \frac{
    G(b^*)\tilde{\psi}(\underline{z}) - G(\underline{z}) \tilde{\psi}(b^*)
    }{
    \tilde{\psi}(\underline{z}) \tilde{\varphi}(b^*) - \tilde{\psi}(b^*) \tilde{\varphi}(\underline{z})
    },
\end{align*}
\endgroup
and in case (\textrm{II}) by 
\begingroup
\allowdisplaybreaks
\begin{align*}
    C_1^* &= \frac{
    \tilde{\varphi}(0) \big( 
    \tilde{\psi}(b^*) [
    G'(\overline{z}) \tilde{\varphi}(\overline{z}) 
    - G(\overline{z}) \tilde{\varphi}'(\overline{z}) ] +
    \tilde{\varphi} (b^*) [
    G(\overline{z}) \tilde{\psi}'(\overline{z}) 
    - G'(\overline{z}) \tilde{\psi}(\overline{z}) 
    ] \big)
    }{
    [
    \tilde{\psi}(b^*) \tilde{\varphi}(0) - \tilde{\psi}(0) \tilde{\varphi}(b^*)
    ] \times 
    [
    \tilde{\psi}(\overline{z}) \tilde{\varphi}'(\overline{z}) -
    \tilde{\psi}'(\overline{z}) \tilde{\varphi}(\overline{z})
    ]
    } \\
    &- \frac{
    \tilde{\varphi}(0) \big( 
    \tilde{\psi}(b^*) [
    G'(\underline{z}) \tilde{\varphi}(\underline{z}) 
    - G(\underline{z}) \tilde{\varphi}'(\underline{z}) ] +
    \tilde{\varphi} (b^*) [
    G(\underline{z}) \tilde{\psi}'(\underline{z}) 
    - G'(\underline{z}) \tilde{\psi}(\underline{z}) 
    ] \big)
    }{
    [
    \tilde{\psi}(b^*) \tilde{\varphi}(0) - \tilde{\psi}(0) \tilde{\varphi}(b^*)
    ] \times 
    [
    \tilde{\psi}(\underline{z}) \tilde{\varphi}'(\underline{z}) -
    \tilde{\psi}'(\underline{z}) \tilde{\varphi}(\underline{z})
    ]
    } \\
    D_1^* &= C_1^* + \frac{
    G'(\underline{z}) \tilde{\varphi}(\underline{z}) - 
    G(\underline{z}) \tilde{\varphi}'(\underline{z})
    }{
    \tilde{\psi}(\underline{z})\tilde{\varphi}'(\underline{z}) - 
    \tilde{\psi}'(\underline{z}) \tilde{\varphi}(\underline{z})
    }, \qquad 
    D_2^* = - C_1^* \frac{\tilde{\psi}(0)}{\tilde{\varphi}(0)} -
    \frac{
    G'(\underline{z})\tilde{\psi}(\underline{z}) - G(\underline{z}) \tilde{\psi}'(\underline{z})
    }{
    \tilde{\psi}(\underline{z}) \tilde{\varphi}'(\underline{z}) - 
    \tilde{\psi}'(\underline{z}) \tilde{\varphi}(\underline{z})
    } \\
    E_1^* &= C_1^* \frac{
    \tilde{\varphi}(b^*) [
    \tilde{\psi}(\overline{z}) \tilde{\varphi}(0) - 
    \tilde{\psi}(0) \tilde{\varphi}(\overline{z})
    ]
    }{
    \tilde{\varphi} (0) [
    \tilde{\psi}(\overline{z}) \tilde{\varphi} (b^*) - 
    \tilde{\psi}(b^*) \tilde{\varphi}(\overline{z})
    ]
    } +
    \frac{
    G(\overline{z}) \tilde{\varphi}(b^*)        
    }{
    \tilde{\psi}(\overline{z})\tilde{\varphi}(b^*) - \tilde{\psi} (b^*) \tilde{\varphi}(\overline{z})
    }\\
    &+ 
    \frac{
    \tilde{\varphi} (b^*) \big( \tilde{\psi} (\overline{z}) [
    G'(\underline{z}) \tilde{\varphi}(\underline{z}) - G(\underline{z}) \tilde{\varphi} '(\underline{z})
    ] - 
    \tilde{\varphi}(\overline{z}) [
    G'(\underline{z}) \tilde{\psi}(\underline(\underline{z}) - 
    G(\underline{z}) \tilde{\psi}'(\underline{z}) 
    ] \big)
    }{
    [
    \tilde{\psi}(\overline{z}) \tilde{\varphi} (b^*) - \tilde{\psi}(b^*) \tilde{\varphi} (\overline{z}) 
    ] \times [
    \tilde{\psi}(\underline{z}) \tilde{\varphi}'(\underline{z}) - 
    \tilde{\psi}'(\underline{z}) \tilde{\varphi}(\underline{z})
    ] 
    } \\
     C_2^* &= - C_1^* \frac{\tilde{\psi}(0)}{\tilde{\varphi}(0)}, \qquad E_2^* = - E_1^* \frac{\tilde{\psi}(b^*)}{\tilde{\varphi}(b^*)}. 
\end{align*}%
\endgroup
We denote the corresponding equilibrium density, that results from plugging in the solutions for the constants $A_i^*, B_i^*, C_i^*, D_i^*, E_i^*, i=1,2$ into \eqref{ECP:Equ:Densf(z)I} -- \eqref{ECP:Equ:Densf(z)II}, as $f^*(z)$. \par
\vspace{0.15cm}
\textit{Step 3.} As a last step, we determine the entry rate $N^*$ via the equilibrium condition \eqref{ECP:Equ:EmissionTarget}. Since $\nu^* (z) = N^* f^*(z)$ is linear in the entry rate, we notice that \eqref{ECP:Equ:EmissionTarget} rewrites as
\begin{align*}
  E_{max} &= \int_0^b e (z;c_p,E_{max}) \nu (dz) = 
  \int_0^b e (z;c_p,E_{max}) N^* f^*(z) dz,
\end{align*}
such that the equilibrium entry rate is given by 
\begin{align}\label{ECP:Equ:EquiEntryRateN}
  N^* = \frac{E_{max}}{\int_0^b e (z;c_p,E_{max}) f^*(z) dz}.
\end{align}
Notice that the integral in the denominator of \eqref{ECP:Equ:EquiEntryRateN} is finite due to Assumption \ref{ECP:Assumption:ProfitFct}.

\section{On Assumption \ref{ECP:Assumption:ProfitFct} for the Case study}\label{ECP:Appendix:ModelsatisfiesAssumption}
    \noindent Clearly, the functions $\pi_1, \pi_2$ and $e$ satisfy the regularity assumptions posed in Assumption \ref{ECP:Assumption:ProfitFct} (i). Moreover, we observe $\pi_i (z) \geq 0$ for all $z\in \mathbb{R}_+$, since $\pi_i (0) = 0$ and $\partial_z \pi_i ( z) >0$. It is straightforward to see that $\partial_{c_p} \pi_1(z,c_p) <0$, as well as $\lim_{c_p \to \infty} \pi_1(z,c_p) = 0.$ Furthermore, we observe $0 \leq \pi_1(z,c_p) \leq K_1 (c_p)$, since $\pi_1 (0, c_p) = 0$, $\partial_z \pi_1(z,c_p) \geq 0$ and 
    \begin{align*}
        \lim_{z \to \infty} \pi_1(z,c_p) = (1- \tau_1) \epsilon \Big( \frac{1-\epsilon}{c_p \lambda} \Big)^{\frac{1-\epsilon}{\epsilon}} =: K_1(c_p).
    \end{align*}
    Similarly, it follows that $0 \leq \pi_2(z) \leq K_2$ for some constant $K_2 >0$. \qed

\singlespacing
\printbibliography

\end{document}